\documentclass[prb,twocolumn,superscriptaddress,floatfix,showkeys]{revtex4}
\usepackage{graphicx}
\usepackage{amsmath,subfigure,epsfig,psfrag}
\usepackage{comment}
\usepackage{commath}
\usepackage{graphicx}
\usepackage{amssymb}
\usepackage{xcolor}
\usepackage{soul}
\usepackage{listings}
\usepackage{rotating}
\usepackage{latexsym}
\usepackage{amsfonts}
\usepackage{amssymb}
\usepackage{wasysym}
\usepackage{xfrac}
\usepackage{mathtools}
\usepackage{multirow}
\usepackage{bbm}
\makeatletter
\renewcommand*\env@matrix[1][\arraystretch]{%
  \edef\arraystretch{#1}%
  \hskip -\arraycolsep
  \let\@ifnextchar\new@ifnextchar
  \array{*\c@MaxMatrixCols c}}
\makeatother

\begin{document}

\title{First-Principles Calculation of the Optical Rotatory Power of Periodic Systems: Modern Theory with Modern Functionals}

\author{Jacques K. Desmarais}
\email{jacqueskontak.desmarais@unito.it}
\affiliation{Dipartimento di Chimica, Universit\`{a} di Torino, via Giuria 5, 10125 Torino, Italy}
\affiliation{Universit\'e de Pau et des Pays de l'Adour, CNRS, IPREM, E2S UPPA, Pau, France}

\author{Bernard Kirtman}
\affiliation{Department of Chemistry and Biochemistry, University of California,
Santa Barbara, California 93106, USA}

\author{Michel R\'erat}
\affiliation{Universit\'e de Pau et des Pays de l'Adour, CNRS, IPREM, E2S UPPA, Pau, France}

\date{\today}
\begin{abstract}
An analysis of orbital magnetization in band insulators is provided. It is shown that a previously proposed electronic orbital angular-momentum operator generalizes the ``modern theory of orbital magnetization'' to include non-local Hamiltonians. Expressions for magnetic transition dipole moments needed for the calculation of optical rotation (OR) and other properties are developed. A variety of issues that arise in this context are critically analyzed. These issues include periodicity of the operators, previously proposed band dispersion terms as well as, if and where needed, evaluation of reciprocal space derivatives of orbital coefficients. Our treatment is used to determine the optical rotatory power of band insulators employing a formulation that accounts for electric dipole - electric quadrupole (DQ), as well as electric dipole-magnetic dipole, contributions.  An implementation in the public \textsc{Crystal} program is validated against a model finite system and applied to the $\alpha$-quartz mineral through linear-response time-dependent density functional theory with a hybrid functional. The latter calculations confirmed the importance of DQ terms. Agreement against experiment was only possible with i) use of a high quality basis set, ii) inclusion of a fraction of non-local Fock exchange, and iii) account of orbital-relaxation terms in the calculation of response functions.
\end{abstract}

\maketitle

\section{Introduction}
\label{sec:intro}

The rotatory power of an optically active material refers to its capacity to rotate the plane of polarization of plane-polarized light. For a non-magnetic medium, the angle of rotation per unit distance $\Phi_u$ of light of wavelength $\lambda$, propogating along the direction $u$ may be expressed in terms of diagonal elements of the optical rotation (OR) tensor $\beta_{u}$:\cite{condon1937theories}
\begin{equation}
\Phi_u = \left( \frac{2 \pi}{\lambda} \right)^2 \frac{4 \pi \beta_{u}}{V}
\end{equation}
where $V$ is the volume per unit cell of the optically active medium. At variance with the notation of a previous article,\cite{rerat2021first} here $u$ is the direction of the light beam.

For finite systems, the theory is well-established based on a multipole expansion of the interaction Hamiltonian $\mathbf{p \cdot A + A \cdot p}$.\cite{hansen1972fully,tinoco2009theoretical} This results in a formula for the OR angle about the $u$ direction that is proportional to the sum of a dynamic magnetic dipole - electric dipole (DD) and an electric dipole - electric quadrupole (DQ) term.\cite{buckingham1971optical,hansen1972fully,tinoco2009theoretical} For samples in solution, orientational averaging then leads to a sum over all directions given by the trace of the DD tensor (the DQ tensor being traceless).\cite{craig1998molecular,buckingham1971optical,hansen1972fully,tinoco2009theoretical} 

For infinite periodic systems (e.g. crystalline solids) the theory of OR or, for that matter, the orbital response to electromagnetic fields in general, is not as straightforward. In the case of electric fields, it was noticed as early as 1962 that integrals of the simple operator $-\mathbf{r}$ over crystalline-orbitals (COs) $\vert \psi_{i,\mathbf{k}} \rangle$:
\begin{equation}
\label{eqn:CO_BF}
\vert \psi_{i,\mathbf{k}} \rangle = e^{\imath \mathbf{k} \cdot \mathbf{r} + \imath \phi_i \left( \mathbf{k} \right)} \vert u_{i,\mathbf{k}} \rangle
\end{equation}
 are undefined, since their value depends on the (arbitrary) choice of the unit cell.\cite{blount1962formalisms} In Eq. (\ref{eqn:CO_BF}) $\vert u_{i,\mathbf{k}} \rangle$ are the cell-periodic Bloch functions and $\phi \left( \mathbf{k} \right)$ is the arbitrary (apart from the constraints provided by periodic boundary conditions) phase of the COs. 

In the same 1962 paper Blount noted that this problem can be avoided by replacing $-\mathbf{r}$ for $-\imath e^{\imath \mathbf{k} \cdot \mathbf{r}} \boldsymbol{\nabla_k} e^{-\imath \mathbf{k} \cdot \mathbf{r}}$, leading to the replacement $-\mathbf{r} \to -\left( \mathbf{r} + \imath \boldsymbol{\nabla_k} \right)$ for periodic systems. The same operator was applied by Otto (1992) as well as Kirtman \textit{et al.} (2000); extended to 2D and 3D by Ferrero \textit{et al.} (2008); and subsequently utilized by many others for the calculation of linear, as well as non-linear optical properties, vibrational spectra, and piezoelectricity, for instance.\cite{rerat2008comparison,kirtman2000extension,bishop2001coupled,mauro1,mauro2,mauro5,RO10,bernasconi,Piezo_rerat_2010,Piezo_rerat_2009,PIEZO_CPHF} This treatment also coincides with King-Smith, Vanderbilt and Resta's ``modern theory of polarization''.\cite{king1993theory,resta1994macroscopic,resta1998quantum,springborg2008analysis}

As for magnetic fields in periodic systems, a suitable theoretical framework was first suggested around the same time by Brown and by Zak in (1964) based on group-theoretic considerations.\cite{zak1964magneticI,zak1964magneticII,brown1964bloch} However, their approach was not formulated in terms of conventional Bloch functions, as used in practical Kohn-Sham density-functional theory (KS-DFT) calculations. Nonetheless, Brown and Zak's analysis provided the starting point for an understanding of the quantum Hall effect (and eventually topological insulators and other topological states of matter) in the work of Thouless and co-workers.\cite{thouless1982quantized,dana1985quantised,kane2005quantum} 

A practical formulation for calculating the orbital magnetization of a band insulator did not occur until much later with Ceresoli, Thonhauser, Vanderbilt and Resta's (CTVR) ``modern theory of orbital magnetization''.\cite{thonhauser2005orbital,ceresoli2006orbital} This approach was soon after extended to Chern insulators and metals by Shi, Vignale, Xiao and Niu (SVXN).\cite{shi2007quantum} SVXN also showed that their treatment was valid in the presence of electron-electron interactions, through a theoretical framework known as spin-current DFT (SCDFT). 

Applications of the CTVR theory have, mostly been limited to tight-binding calculations or calculations with local or generalized gradient approximations, but not the hybrid functionals of generalized Kohn-Sham (GKS) theory, and without orbital relaxation effects.\cite{xiao2010berry,essin2009magnetoelectric,ceresoli2006orbital,resta2010electrical,malashevich2010theory,souza2008dichroic,malashevich2012full,ceresoli2007orbital,thonhauser2009converse,thonhauser2009nmr,seidl1996generalized} Indeed, some of our recent work has shown that hybrid functionals may be particularly important within the framework of SCDFT, by providing a non-local dependence of the exchange-correlation (xc) functional on current densities, as practically demonstrated in the case of Weyl semimetals and $\mathbb{Z}_2$ topological insulators.\cite{comaskey2022spin,desmarais2019fundamental,desmarais2019spin,desmarais2021spin,desmarais2021perturbationII,desmarais2020adiabatic,bodo2022spin,desmarais2020spin,desmarais2020development} Here we show that a generalization of CTVR to include non-local functionals coincides with a previous treatment of orbital response to magnetic fields by Springborg Molayem and Kirtman (SMK).\cite{springborg2017electronic}

The CTVR theory pertains to expectation values for calculation of the orbital magnetization (a first order property). For higher orders (e.g. OR, a second-order property), we require transition moments between ground and excited states. In this paper we resolve a variety of issues that remain concerning the theory of OR.

Remaining issues include i) the agreement (or lack thereof) versus large finite systems, ii) the possibility of missing terms proportional to the derivative in reciprocal space of coefficients of the perturbed wave function, iii) the possibility of missing ``band dispersion'' terms iv) a strategy for the effective treatment of quasi-degeneracies v) the importance of DQ terms for calculations in the solid state. Questions concerning items i), ii), and v) are explicitly raised in Ref. \onlinecite{balduf2022derivation}; item iii) occurs as a result of the treatment in Ref. \onlinecite{wang2022optical}; and item iv) is a key computational issue that occurs in evaluating matrix elements of $\boldsymbol{\nabla_k}$.

The paper is organized as follows: In section \ref{sec:form_wannier} we develop the formalism for the orbital magnetization from a Wannier function perspective. This analysis shows how the SMK ``angular momentum'' generalizes CTVR's ``modern theory'' to non-local functionals. In section \ref{sec:form_pert} we show how the same formalism for orbital magnetization may alternatively be developed from a perturbation theory perspective. This provides a convenient route for developing expressions for transition moments (and, thus, higher order properties) and provides answers regarding items i), ii) and v). Technical aspects of the formal developments are discussed in appendices, including particularly Appendix \ref{app:band_disp} regarding iii). 
\section{Formalism: Wannier Function Perspective}
\label{sec:form_wannier}

\subsection{Review of Modern Theory for Band Insulators}
\label{sec:modern}

With infinite periodic systems, the eigenfunctions $\vert \psi_{i,\mathbf{k}} \rangle$  of the single particle Hamiltonian:
\begin{equation}
\label{eqn:Hpsi_epsi}
\hat{H} \vert \psi_{i,\mathbf{k}} \rangle = \epsilon_{i,\mathbf{k}} \vert \psi_{i,\mathbf{k}} \rangle
\end{equation}
are crystalline orbitals (COs), which may be written in Bloch form as in Eq. (\ref{eqn:CO_BF}).

A key quantity in the discussion of magnetic properties of periodic systems is the Chern (vector) invariant $\mathbf{C}$:\cite{thouless1982quantized}
\begin{equation}
\label{eqn:chern}
\mathbf{C}= \frac{\imath}{2 \pi} \int_\text{BZ} d \mathbf{k} \ \sum_i^\text{occ} \langle \boldsymbol{\nabla_k} u_{i,\mathbf{k}} \vert \wedge \vert \boldsymbol{\nabla_k} u_{i,\mathbf{k}} \rangle
\end{equation}
In Eq. (\ref{eqn:chern}) and elsewhere the superscript occ refers to the occupied manifold while the subscript BZ indicates integration over the first Brillouin zone. As will be discussed below, we are concerned here with systems where $\mathbf{C}$ vanishes in the absence of a magnetic field:
\begin{equation}
\label{eqn:chern_zero}
\mathbf{C}=\mathbf{0}
\end{equation}

In the case of insulating systems with a vanishing  $\mathbf{C}$, a unitary transformation of the BFs provides localized Wannier functions (WFs) $\vert w_{i,\mathbf{g}} \rangle$:
\begin{subequations}
\label{eqn:bloch_wann}
\begin{equation}
\label{eqn:wannier}
\vert w_{i,\mathbf{g}} \rangle = \Omega \int_\text{BZ} d \mathbf{k} \ e^{\imath \mathbf{k} \cdot \left( \mathbf{r} - \mathbf{g} \right)} \vert u_{i,\mathbf{k}} \rangle
\end{equation}
with the inverse transform given by:
\begin{equation}
\label{eqn:bloch_w}
\vert u_{i,\mathbf{k}} \rangle = \sum_{\mathbf{g}} e^{-\imath \mathbf{k} \cdot \left( \mathbf{r} - \mathbf{g} \right)} \vert w_{i,\mathbf{g}} \rangle
\end{equation}
\end{subequations}
where $\Omega$ is the volume of the FBZ and $\mathbf{g}$ is a direct lattice vector (with the corresponding sum running over the infinite set). The transformation of Eq. (\ref{eqn:bloch_wann}) is of course, like the orbitals themselves, only determined up to an arbitrary phase $e^{\imath \phi_i^\prime \left( \mathbf{k} \right)}$. For the moment we set $\phi_i^\prime \left( \mathbf{k} \right)=0$, its effect being discussed further on.

In the case of band-insulators, the density matrix has exponential decay in direct space,\cite{CRYSTAL88_1} and it is possible to represent the orbitals in well-localized form, as per Eq. (\ref{eqn:wannier}). This procedure is also applicable to $\mathbb{Z}_2$ topological insulators.\cite{soluyanov2011wannier} On the other hand, for metals, the power-law decay of the density matrix means that the localization procedure diverges.\cite{resta2006kohn} A similar behavior occurs for Chern insulators, due to the existence of conducting chiral edge states.\cite{thonhauser2006insulator}

For ``Wannier representable'' band (or $\mathbb{Z}_2$ topological) insulators, CTVR write the orbital magnetization of a large, finite sample of $N_c$ cells cut from the bulk with $N_b$ orbitals as:\cite{ceresoli2006orbital}
\begin{equation}
\label{eqn:orb_mag_w}
\mathbf{M} = -\frac{1}{2c \Omega N_c} \sum_i^{N_b} \langle w_{i} \vert \mathbf{r} \wedge \imath \left[ H, \mathbf{r} \right]  \vert w_{i}  \rangle
\end{equation}
where $c$ is the speed of light and the index $\mathbf{g}$ has been dropped for general large, finite samples. Eq. (\ref{eqn:orb_mag_w}) is justified and computationally convenient for local Hamiltonians, with matrix elements of the velocity operator $\hat{\mathbf{v}}$ that read:
\begin{equation}
\label{eqn:velocity}
\langle w_{i} \vert  \boldsymbol{\nabla_r} / \imath \vert w_{i}  \rangle = \langle w_{i}  \vert  \imath \left[ H, \mathbf{r} \right]  \vert w_{i}  \rangle
\end{equation}
Indeed, substituting Eq. (\ref{eqn:velocity}) into Eq. (\ref{eqn:orb_mag_w}), $\mathbf{M}$ reduces to the standard first-order perturbation theory expression, in terms of a matrix elements of the electronic angular-momentum operator $\hat{\mathbf{L}}=\mathbf{r} \wedge \boldsymbol{\nabla_r} / \imath$. At this point, we note that the situation is different, for calculations with a non-local GKS Hamiltonian since the replacement:
\begin{equation}
\label{eqn:rep_nabla}
- \left[ H, \mathbf{r} \right] \to \boldsymbol{\nabla_r}
\end{equation}
fails to account for the fact that $\mathbf{r}$ does not commute with the non-local terms. Of course, the non-commutativity may be corrected for hybrid functionals, for instance, by an explicit calculation of the commutator $\left[ H_\text{Fock}, \mathbf{r}\right]$ with the Fock operator $H_\text{Fock}$. However, this is impractical for actual calculations as it involves the evaluation of electron-electron integrals of increased complexity.

In the thermodynamic limit (i.e. $N_s / N_b \to 0$, with $N_b$ the number of WFs $w_{i}$ centered in the interior of the sample and $N_s$ the number centered at the surface), using translational symmetry, the contribution to Eq. (\ref{eqn:orb_mag_w}) from the bulk WFs becomes:\cite{ceresoli2006orbital}
\begin{equation}
\label{eqn:M_bulk}
\mathbf{M}_\text{LC} = -\frac{1}{2c \Omega} \sum_i^\text{occ} \langle w_{i,\mathbf{0}} \vert \mathbf{r} \wedge \imath \left[ H, \mathbf{r} \right]  \vert w_{i,\mathbf{0}}  \rangle
\end{equation}
On the other hand, the contribution to Eq. (\ref{eqn:orb_mag_w}) from the $N_s$ surface orbitals reads:\cite{ceresoli2006orbital}
\begin{eqnarray}
\label{eqn:M_surf}
\mathbf{M}_\text{IC} &=& -\frac{1}{2c \Omega N_c} \Big[ \sum_s^{N_s} \langle w_{s} \vert \left( \mathbf{r} - \mathbf{r}_s \right) \wedge \imath \left[ H, \mathbf{r} \right]  \vert w_{s}  \rangle \nonumber \\
&+& \mathbf{r}_s \wedge  \langle w_{s} \vert \imath \left[ H, \mathbf{r} \right]  \vert w_{s}  \rangle \Big]
\end{eqnarray}

where, we have used the shorthand notation $\mathbf{r}_s=\langle w_{s} \vert \mathbf{r} \vert w_{s} \rangle$ to denote the centers of the surface WFs. Indeed, in the thermodynamic limit, even though the first term vanishes, $\mathbf{r}_s$ is large and a non-vanishing contribution from the second term in Eq. (\ref{eqn:M_surf}) remains.\cite{ceresoli2006orbital} Then, Eq. (\ref{eqn:M_bulk}) is related to currents circulating in the bulk of the solid (the so-called local-circulation, LC, term), while Eq. (\ref{eqn:M_surf}) originates from current circulating at the surface of the solid (the itinerant-circulation, IC, term). Remarkably, however, both contributions may be rewritten solely in terms of WFs centered in the bulk. 

In reciprocal space, the final expressions are, for the LC term:\cite{ceresoli2006orbital}
\begin{subequations}
\begin{equation}
\label{eqn:MLC}
\mathbf{M}_\text{LC} = \frac{1}{2c} \Re \int_\text{BZ} d \mathbf{k} \sum_i^\text{occ} \langle \boldsymbol{\nabla_k} u_{i,\mathbf{k}} \vert \wedge H_{\mathbf{k}} / \imath  \vert \boldsymbol{\nabla_k} u_{i,\mathbf{k}} \rangle
\end{equation}
and for the IC term:\cite{ceresoli2006orbital}
\begin{eqnarray}
\label{eqn:MIC}
\mathbf{M}_\text{IC} &=& \frac{1}{2c} \Re \int_\text{BZ} d \mathbf{k} \sum_l^\text{all}  \sum_i^\text{occ}   \langle \boldsymbol{\nabla_k} u_{i,\mathbf{k}} \vert \wedge \vert \boldsymbol{\nabla_k} u_{l,\mathbf{k}} \rangle \nonumber \\
&\times& \langle u_{l,\mathbf{k}} \vert H_{\mathbf{k}} / \imath \vert u_{i,\mathbf{k}} \rangle
\end{eqnarray}
\end{subequations}
where:
\begin{equation}
\label{eqn:HkH}
H_{\mathbf{k}} = e^{- \imath \mathbf{k} \cdot \mathbf{r}} H e^{\imath \mathbf{k} \cdot \mathbf{r}}
\end{equation}
The total orbital-magnetization is the sum of the LC and IC contributions:\cite{ceresoli2006orbital}
\begin{equation}
\label{eqn:M}
\mathbf{M} = \mathbf{M}_\text{LC} + \mathbf{M}_\text{IC}
\end{equation}
In deriving Eq. (\ref{eqn:MIC}), CTVR insert $\mathbbm{1} = \sum_l^\text{all} \vert w_{l}  \rangle \langle w_{l} \vert$ into Eq. (\ref{eqn:M_surf}), yielding the relation:
\begin{equation}
\label{eqn:cbs}
\langle w_{s} \vert \imath \left[ H, \mathbf{r} \right]  \vert w_{s}  \rangle = 2 \Re \sum_l^\text{all} \langle w_{s} \vert \mathbf{r} \vert w_{l}  \rangle \langle w_{l} \vert  H / \imath \vert w_{s}  \rangle
\end{equation}
Here we note that for calculations in a finite basis set, it is then necessary to remove the identity, from resulting expressions. In other words, for finite basis set calculations it is necessary to make the replacement: 
\begin{equation}
\label{eqn:rep_identity}
\sum_l^\text{all} \vert w_{l}  \rangle \langle w_{l} \vert \to \mathbbm{1}
\end{equation}
in Eq. (\ref{eqn:MIC}) for $\mathbf{M}_\text{IC}$. 

In the following, we develop expressions that are generally valid for non-local Hamiltonians and a finite basis set, in a manner that achieves the replacements warranted by Eqs. (\ref{eqn:rep_nabla}) and (\ref{eqn:rep_identity}). The result shows a connection with a previously proposed treatment by SMK.\cite{springborg2017electronic}

\subsection{Expressions for non-local Hamiltonians and a finite basis}

\subsubsection{Local Circulation Term}

For further development, it proves useful to develop the derivatives of Eq. (\ref{eqn:bloch_w}):
\begin{subequations}
\begin{equation}
\label{eqn:dkbloch_w}
\boldsymbol{\nabla_k} \vert u_{i,\mathbf{k}} \rangle = - \imath \sum_{\mathbf{g}} e^{-\imath \mathbf{k} \cdot \left( \mathbf{r} - \mathbf{g} \right)} \left( \mathbf{r} - \mathbf{g} \right)  \vert w_{i,\mathbf{g}} \rangle
\end{equation}
and:
\begin{equation}
\label{eqn:drbloch_w}
\boldsymbol{\nabla_r} \vert u_{i,\mathbf{k}} \rangle = -\imath \mathbf{k}  \vert u_{i,\mathbf{k}} \rangle +  \sum_\mathbf{g} e^{-\imath \mathbf{k} \cdot \left( \mathbf{r} - \mathbf{g} \right)} \boldsymbol{\nabla_r} \vert w_{i,\mathbf{g}} \rangle
\end{equation}
\end{subequations}
Furthermore, inserting Eq. (\ref{eqn:HkH}) into Eq. (\ref{eqn:dkbloch_w}) provides:
\begin{equation}
\label{eqn:Hkdkbloch_w}
H_\mathbf{k} \boldsymbol{\nabla_k} \vert u_{i,\mathbf{k}} \rangle = - \imath \sum_{\mathbf{g}} e^{-\imath \mathbf{k} \cdot \left( \mathbf{r} - \mathbf{g} \right)} H \left( \mathbf{r} - \mathbf{g} \right)  \vert w_{i,\mathbf{g}} \rangle
\end{equation}
In Appendix  \ref{app:Hgradk} we use Eqs. (\ref{eqn:dkbloch_w})-(\ref{eqn:Hkdkbloch_w}), the zero-Chern invariant condition of Eq. (\ref{eqn:chern_zero}), and the replacement of Eq. (\ref{eqn:rep_nabla}), required for non-local Hamiltonians to develop the following relation:
\begin{eqnarray}
\label{eqn:u_Hgradk1}
\wedge H_\mathbf{k} / \imath \vert \boldsymbol{\nabla_k} u_{i,\mathbf{k}} \rangle &\to&  \wedge \sum_\mathbf{g} e^{-i \mathbf{k} \cdot \left( \mathbf{r} - \mathbf{g} \right)} \boldsymbol{\nabla_r} \vert w_{i,\mathbf{g}} \rangle \nonumber \\
&=& \wedge \boldsymbol{\nabla_r} \vert u_{i,\mathbf{k}} \rangle + \wedge \imath \mathbf{k} \vert u_{i,\mathbf{k}} \rangle
\end{eqnarray}
Finally, inserting Eq. (\ref{eqn:u_Hgradk1}) into Eq. (\ref{eqn:MLC}) yields:
\begin{eqnarray}
\label{eqn:MLC2}
\mathbf{M}_\text{LC} &=& \frac{1}{2c} \Re \int_{\text{BZ}} d \mathbf{k} \sum_i^\text{occ}  \big[ \langle \boldsymbol{\nabla_k} u_{i,\mathbf{k}} \vert \wedge \boldsymbol{\nabla_r} \vert u_{i,\mathbf{k}} \rangle \nonumber \\
&+& \langle \boldsymbol{\nabla_k} u_{i,\mathbf{k}} \vert \wedge \imath \mathbf{k} \vert u_{i,\mathbf{k}} \rangle \big]
\end{eqnarray}
Then, applying $\boldsymbol{\nabla_k}$ to Eq. (\ref{eqn:CO_BF}), we obtain a representation of the position operator for a periodic system due to Blount:\cite{blount1962formalisms}
\begin{equation}
\label{eqn:Blount}
\mathbf{r} \vert \psi_{i,\mathbf{k}} \rangle = \imath e^{\imath \mathbf{k} \cdot \mathbf{r}} \boldsymbol{\nabla_k} \vert u_{i,\mathbf{k}} \rangle - \imath \boldsymbol{\nabla_k} \vert \psi_{i,\mathbf{k}} \rangle
\end{equation}
Inserting Eq. (\ref{eqn:Blount}) into Eq. (\ref{eqn:MLC2}) provides:
\begin{equation}
\label{eqn:MLC3}
\mathbf{M}_\text{LC} = - \frac{1}{2c} \Re \int_{\text{BZ}} d \mathbf{k}  \sum_i^\text{occ} \langle \psi_{i,\mathbf{k}} \vert \left( \mathbf{r} + \imath \boldsymbol{\nabla_k} \right)^\dagger \wedge \boldsymbol{\nabla_r} / \imath \vert \psi_{i,\mathbf{k}} \rangle
\end{equation}

\subsubsection{Itinerant Circulation Term}

Returning to Eq. (\ref{eqn:MIC}) and using the Hermiticity of $H_{\mathbf{k}}$, we obtain:
\begin{eqnarray}
\label{eqn:MIC1}
\mathbf{M}_\text{IC} &=& \frac{1}{2c} \Re \int_\text{BZ} d \mathbf{k} \sum_l^\text{all}  \sum_i^\text{occ}   \langle \boldsymbol{\nabla_k} u_{i,\mathbf{k}} \vert \wedge \vert \boldsymbol{\nabla_k} u_{l,\mathbf{k}} \rangle \nonumber \\
&\times& \langle u_{l,\mathbf{k}} \vert H_{\mathbf{k}} / \imath \vert u_{i,\mathbf{k}} \rangle \nonumber \\
&=& \frac{1}{2c} \Re \int_\text{BZ} d \mathbf{k} \sum_l^\text{all}  \sum_i^\text{occ} \langle u_{i,\mathbf{k}} \vert H_{\mathbf{k}} / \imath \vert u_{l,\mathbf{k}} \rangle \nonumber \\
&\times& \langle \boldsymbol{\nabla_k} u_{l,\mathbf{k}} \vert \wedge \vert \boldsymbol{\nabla_k} u_{i,\mathbf{k}} \rangle
\end{eqnarray}
The combination of Eq. (\ref{eqn:rep_identity}) with Eq. (\ref{eqn:MIC1}) gives:
\begin{equation}
\label{eqn:MIC2}
\mathbf{M}_\text{IC} =  \frac{1}{2c} \Re \int_\text{BZ} d \mathbf{k} \sum_i^\text{occ} \langle u_{i,\mathbf{k}} \vert H_{\mathbf{k}} / \imath  \boldsymbol{\nabla_k}^\dagger \wedge  \boldsymbol{\nabla_k} \vert u_{i,\mathbf{k}} \rangle
\end{equation}
By utilizing the conjugate-transpose of Eq. (\ref{eqn:u_Hgradk1}) along with  Eq. (\ref{eqn:Blount}), Eq. (\ref{eqn:MIC2}) leads to:
\begin{equation}
\label{eqn:MIC3}
\mathbf{M}_\text{IC} =  \frac{1}{2c} \Re \int_\text{BZ} d \mathbf{k} \sum_i^\text{occ} \langle \psi_{i,\mathbf{k}} \vert \boldsymbol{\nabla_r}^\dagger / \imath \wedge \left( \mathbf{r} + \imath \boldsymbol{\nabla_k} \right) \vert \psi_{i,\mathbf{k}} \rangle
\end{equation}

\subsubsection{SMK ``Angular-Momentum'' Operator}
From the combination of Eqs. (\ref{eqn:M}), (\ref{eqn:MLC3}) and (\ref{eqn:MIC3}) we obtain an expression for the orbital magnetization in terms of the SMK ``Hermitized angular-momentum'' operator:
\begin{subequations}
\label{eqn:M_SMK}
\begin{equation}
\mathbf{M} = - \frac{1}{2c} \int_\text{BZ} d \mathbf{k} \sum_i^\text{occ} \langle \psi_{i,\mathbf{k}} \vert \frac{\hat{\boldsymbol{\Lambda}} + \hat{\boldsymbol{\Lambda}}^\dagger}{2} \vert \psi_{i,\mathbf{k}} \rangle
\end{equation}
where:
\begin{equation}
\label{eqn:Lambda_op}
\hat{\boldsymbol{\Lambda}} =  \left( \mathbf{r} + \imath \boldsymbol{\nabla_k} \right) \wedge \boldsymbol{\nabla_r} / \imath
\end{equation}
\end{subequations}
The SMK theory, thus corresponds to a generalization of CTVR's modern theory of orbital magnetization to calculations with (generally) non-local Hamiltonians in a finite basis set.

\section{Formalism: Perturbation Theory Perspective}
\label{sec:form_pert}

The result obtained in section \ref{sec:form_wannier} (for the magnetization in terms of a ``Hermitized angular momentum'' operator) that was derived using Wannier functions may alternatively be developed through perturbation theory (PT). The latter approach is similar to the derivation presented by SVXN in generalizing the CTVR treatment to metals and Chern insulators.\cite{shi2007quantum} This approach is also more convenient for extension to higher-order properties. 

We begin by considering a transition matrix element of the first-order interaction Hamiltonian for the orbital response to a magnetic field which, in the CO basis, is:
\begin{equation}
\label{eqn:H1}
-\frac{\imath}{c} \langle \psi_{i,\mathbf{k}} \vert \boldsymbol{\nabla_r} \cdot \mathbf{A} + \mathbf{A} \cdot \boldsymbol{\nabla_r} \vert \psi_{a,\mathbf{k}^\prime}  \rangle 
\end{equation}
In Eq. (\ref{eqn:H1}) $\mathbf{A}$ is the magnetic vector potential. In the Coulomb gauge (for which $\boldsymbol{\nabla_r} \cdot \mathbf{A} = \mathbf{0}$):
\begin{equation}
\label{eqn:AB_final}
\mathbf{A}  = \imath \mathbf{B}_0 \wedge \frac{\mathbf{q}}{q^2} e^{\imath \left( \mathbf{q} \cdot \mathbf{r} - \omega t \right)}
\end{equation}

where $\mathbf{B}_0$ is the (constant) amplitude of the magnetic field with wavevector $\mathbf{q}$ and frequency $\omega$ ($q=\vert \mathbf{q} \vert$). Then, for a spatially oscillating magnetostatic field $\mathbf{B}=\left[ B_x,0,0 \right]^T$ propagating in a direction orthogonal to $x$ (i.e. transversal wave, $q_x=0$):
\begin{equation}
\label{eqn:A_final}
\mathbf{A} \left( \mathbf{r} \right)  = \imath B_{0x} 
\begin{pmatrix}
0 \\
 - q_z \ e^{\imath \left( q_y r_y + q_z r_z\right)} / q^2\\
q_y \ e^{\imath \left( q_y r_y + q_z r_z\right)} / q^2
\end{pmatrix}
\end{equation}
and $q^2=q_y^2+q_z^2$.

Inserting Eq. (\ref{eqn:A_final}) into Eq. (\ref{eqn:H1}) provides:
\begin{eqnarray}
\label{eqn:lambda_herm1}
\frac{B_{0x}}{c q^2} \langle \psi_{i,\mathbf{k}} \vert  e^{\imath \left( q_y r_y + q_z r_z\right)}  \left( q_y \nabla_z - q_z \nabla_y \right) \vert \psi_{a,\mathbf{k}^\prime}  \rangle
\end{eqnarray}
In Appendix \ref{app:mom}, we show that this integral is null, except when:
\begin{equation}
\label{eqn:kkq}
\mathbf{k}-\mathbf{k}^\prime=\mathbf{q}
\end{equation}
We now expand $e^{\imath \left( q_y r_y + q_z r_z\right)}$, $\vert \psi_{i,\mathbf{k}} \rangle$ and $\vert \psi_{j,\mathbf{k}^\prime} \rangle$ to first order in $\mathbf{q}$, around the midpoint $\bar{\mathbf{k}}=\left( \mathbf{k}+\mathbf{k}^\prime \right) / 2$, using also Eq. (\ref{eqn:kkq}), yielding: 
\begin{eqnarray}
\label{eqn:lambda_herm2}
 \frac{B_{0x}}{c q^2} \langle \psi_{i,\bar{\mathbf{k}}} + \left( \frac{q_y}{2} \nabla_{\bar{k}_y} + \frac{q_z}{2} \nabla_{\bar{k}_z} \right) \psi_{i,\bar{\mathbf{k}}} \vert \nonumber \\
\left[ 1 + \imath \left( q_y r_y + q_z r_z\right) \right] \left( q_y \nabla_z - q_z \nabla_y \right) \nonumber \\
\vert \psi_{a,\bar{\mathbf{k}}} - \left( \frac{q_y}{2} \nabla_{\bar{k}_y} + \frac{q_z}{2} \nabla_{\bar{k}_z} \right) \psi_{a,\bar{\mathbf{k}}} \rangle  
\end{eqnarray}
For a spatially homogeneous field we take the limit $\mathbf{q} \to \mathbf{0}$ of Eq. (\ref{eqn:lambda_herm2}). Then, we write $q_y=q \cos \theta$ and $q_z=q \sin \theta$, and average over $\theta$, which gives: 
\begin{eqnarray}
\frac{B_{0x}}{2 c } \langle \psi_{i,\bar{\mathbf{k}}}  \vert \imath r_y \nabla_z - \imath r_z \nabla_y + \frac{1}{2} \nabla_{\bar{k}_y}^\dagger \nabla_z - \frac{1}{2} \nabla_{\bar{k}_z}^\dagger \nabla_y \nonumber \\
- \frac{1}{2} \nabla_{\bar{k}_y} \nabla_z + \frac{1}{2} \nabla_{\bar{k}_z} \nabla_y \vert \psi_{a,\bar{\mathbf{k}}}  \rangle
\end{eqnarray}

where we have used $\langle \cos^2 \theta \rangle = \langle \sin^2 \theta \rangle = \frac{1}{2}$, as well as $\langle \cos \theta \rangle = \langle \sin \theta \rangle = 0$. Repeating the process for $y$ and $z$-directed fields we get:
\begin{equation}
\label{eqn:lambda_herm3}
- \frac{\mathbf{B}_{0}}{2 c } \cdot \langle \psi_{i,\mathbf{k}}  \vert \left[ \mathbf{r} + \frac{\imath}{2} \left( \boldsymbol{\nabla}_\mathbf{k} - \boldsymbol{\nabla}_\mathbf{k}^\dagger \right) \right] \wedge  \frac{\boldsymbol{\nabla_r}}{\imath} \vert \psi_{a,\mathbf{k}}  \rangle 
\end{equation}
Finally, employing the Hermiticity of $\mathbf{r}$ and the anti-Hermiticity of $\boldsymbol{\nabla_r}$, Eq. (\ref{eqn:lambda_herm3}) may be written in terms of the Hermitized SMK operator of Eq. (\ref{eqn:Lambda_op}):
\begin{equation}
\label{eqn:lambda_herm4}
- \frac{\mathbf{B}_{0}}{2 c } \cdot \langle \psi_{i,\mathbf{k}}  \vert \frac{\hat{\boldsymbol{\Lambda}} + \hat{\boldsymbol{\Lambda}}^\dagger}{2} \vert \psi_{a,\mathbf{k}}  \rangle 
\end{equation}
Eq. (\ref{eqn:lambda_herm4}) is the analogue (for transition moments) of Eq. (\ref{eqn:M_SMK}) of section \ref{sec:form_wannier} for the orbital magnetization.

\section{Effect of the Orbital Phases and Gauge-Origin Invariance}
\label{sec:gauge}
It was mentionned in section \ref{sec:modern}, that the complex Bloch orbitals and WFs are only determined up to an arbitrary phase $e^{\imath \phi_i \left( \mathbf{k} \right)}$. From Eq. (\ref{eqn:Lambda_op}), the effect on the orbital magnetization of changing the phase of the Bloch orbitals is to make the replacement:
\begin{equation} 
\label{eqn:phase}
\mathbf{r} \to \mathbf{r} - \boldsymbol{\nabla_k} \phi_i \left( \mathbf{k} \right)
\end{equation}
in matrix elements of $\hat{\boldsymbol{\Lambda}}$. Therefore, changing the phase of $\vert \psi_{i,\mathbf{k}} \rangle$ has exactly the same effect as changing the gauge-origin by an amount $- \boldsymbol{\nabla_k} \phi_i \left( \mathbf{k} \right)$ for the matrix element involving orbital $\vert \psi_{i,\mathbf{k}} \rangle$. Eq. (\ref{eqn:phase}) has already been pointed out by SMK. The particular choice of the phase, then, becomes irrelevant for gauge-origin invariant calculations (for instance, at the complete basis set, CBS, limit, see Eq. (\ref{eqn:m_gauge}) below). In our calculations, we set the term $- \boldsymbol{\nabla_k} \phi_i \left( \mathbf{k} \right)$ to zero. Generally speaking, an approximate calculation may depend on the gauge-origin, and, by extension on the choice of the phase. In the particular case of OR, R{\'e}rat and Kirtman have shown that the trace and optic axis component of the OR tensor are gauge-origin invariant if the velocity operator is used for the electric field Hamiltonian.\cite{rerat2021first} 

Another criterion of relevance for Eq. (\ref{eqn:M_SMK}) is the behaviour of the ``angular-momentum'' operator under a lattice translation from $\hat{\boldsymbol{\Lambda}} \left( \mathbf{r} \right)$ to $\hat{\boldsymbol{\Lambda}} \left( \mathbf{r}+\mathbf{g} \right)$. A periodic operator $\hat{O}$ satisfies $\hat{O} \left( \mathbf{r} \right) = \hat{O} \left( \mathbf{r}+\mathbf{g} \right)$. We can work out the effect of translation by a lattice vector $\mathbf{g}$ by returning to Eq. (\ref{eqn:M_SMK}), from which we find:
\begin{equation}
- \frac{1}{4c} \int_\text{BZ} d \mathbf{k} \sum_i^\text{occ} \langle \psi_{i,\mathbf{k}} \vert \left( \mathbf{r} + \mathbf{g} + \imath \boldsymbol{\nabla_k} \right) \wedge \boldsymbol{\nabla_r} / \imath \vert \psi_{i,\mathbf{k}} \rangle + \text{H.c.}
\end{equation}
The translation therefore leads to the (generally) non-vanishing contribution:
\begin{equation}
- \frac{1}{4c} \int_\text{BZ} d \mathbf{k} \sum_i^\text{occ} \mathbf{g} \wedge \langle \psi_{i,\mathbf{k}} \vert \boldsymbol{\nabla_r} / \imath \vert \psi_{i,\mathbf{k}} \rangle + \text{H.c.}
\end{equation}
and, in the general case, we obtain $\hat{\boldsymbol{\Lambda}} \left( \mathbf{r} \right) \ne \hat{\boldsymbol{\Lambda}} \left( \mathbf{r}+\mathbf{g} \right)$  Of course, the translation is equivalent to changing the gauge-origin by $\mathbf{g}$. Then, just like in the case of gauge-origin, $\hat{\boldsymbol{\Lambda}}$ becomes periodic if the orbitals are exact (i.e. at the CBS limit). Indeed, at the CBS limit, we may insert Eq. (\ref{eqn:rep_identity}), as well as Eq. (\ref{eqn:rep_nabla}) into Eq. (\ref{eqn:M_SMK}) yielding:
\begin{eqnarray}
\label{eqn:m_gauge}
\mathbf{M} &=& - \frac{1}{4c} \int_\text{BZ} d \mathbf{k} \sum_l^\text{all} \sum_i^\text{occ} \langle \psi_{i,\mathbf{k}} \vert \left( \mathbf{r} + \imath \boldsymbol{\nabla_k} \right) \vert \psi_{l,\mathbf{k}} \rangle \nonumber \\
&\wedge& \langle \psi_{l,\mathbf{k}}  \vert  \boldsymbol{\nabla_r} / \imath \vert \psi_{i,\mathbf{k}} \rangle + \text{H.c.} \nonumber \\
&=&  \frac{1}{4c}  \int_\text{BZ} d \mathbf{k} \sum_l^\text{all} \sum_i^\text{occ} \frac{\langle \psi_{i,\mathbf{k}} \vert \boldsymbol{\nabla_r} \vert \psi_{l,\mathbf{k}} \rangle}{ \epsilon_{i,\mathbf{k}} - \epsilon_{l,\mathbf{k}}  } \nonumber \\
&\wedge& \langle \psi_{l,\mathbf{k}}  \vert  \boldsymbol{\nabla_r} / \imath \vert \psi_{i,\mathbf{k}} \rangle + \text{H.c.} 
\end{eqnarray}
which is obviously both periodic and gauge-origin invariant, and is also independent on the choice of $- \boldsymbol{\nabla_k} \phi_i \left( \mathbf{k} \right)$. 

Because $\hat{\boldsymbol{\Lambda}}$ is only periodic at the CBS limit, in practice, exact comparison against large finite systems can only be expected with a large basis set. Moreover, the usual reciprocity between sampling of reciprocal space (i.e. number of $\mathbf{k}$ points to sample the FBZ) and direct-space (i.e. size of the supercell expansion) is only verified at the CBS limit. In section \ref{sec:res} we present numerical results that explicitly demonstrate this behaviour.

In the case of finite systems, it has very recently been shown that a finite basis set of London atomic orbitals yields the same translational behaviour as the exact orbitals.\cite{peters2022magnetic} Whether a similar relationship can be developed for infinite, periodic systems remains to be demonstrated.

\section{Extension to Optical Rotation}

\subsection{Operators for the Rotatory Strengths}

For OR, following Stephen,\cite{stephen1958double} Tinoco,\cite{tinoco2009theoretical} Snir and Schellman,\cite{snir1973optical} Hansen and Avery \cite{hansen1972fully} and subsequent work \cite{pedersen1995ab,pedersen1999coupled} it is convenient to describe the interaction of the system with circularly polarized light using the electromagnetic $\mathbf{A}$-gauge (vanishing scalar potential) leading to the interaction Hamiltonian:
\begin{equation}
\left( \exp \left[ \imath \mathbf{q} \cdot \mathbf{r} \right] \mathbf{A}_0 \left( \mathbf{q} \right) + \text{c.c.} \right) \cdot \mathbf{p} +  \mathbf{p} \cdot \left( \mathbf{A}_0 \left( \mathbf{q} \right) \exp \left[ \imath \mathbf{q} \cdot \mathbf{r} \right] + \text{c.c.} \right)
\end{equation}
In the following, we generalize the theory of Stephen and co-workers to periodic systems. This allows us to extend a previous treatment,\cite{rerat2021first} to include not only DD, but also DQ contributions to the OR tensor.

We assume here that $\mathbf{A}_0 \left( \mathbf{q} \right)$ is sufficiently slowly varying so that only the leading terms in $\exp \left[ \pm \imath \mathbf{q} \cdot \mathbf{r} \right] \approx 1 \pm \imath \mathbf{q} \cdot \mathbf{r}$ will contribute to the response. Choosing $\mathbf{q}= \hat{\mathbf{z}} q$, differences in absorption of left- and right-circularly polarized light, lead to rotatory strengths proportional to:\cite{hansen1972fully,stephen1958double}
\begin{equation}
\label{eqn:H_em_matele}
\hat{\mathbf{z}}  \cdot \langle \psi_{i,\mathbf{k}} \vert \exp \left[ -\imath \mathbf{q} \cdot \mathbf{r} \right] \mathbf{p} \vert \psi_{a,\mathbf{k}^\prime} \rangle \wedge \langle \psi_{a,\mathbf{k}^\prime} \vert \exp \left[ \imath \mathbf{q} \cdot \mathbf{r} \right] \mathbf{p} \vert \psi_{i,\mathbf{k}} \rangle 
\end{equation}
including effects from multipoles of all orders. Expansion of Eq. (\ref{eqn:H_em_matele}) to first order in $\mathbf{q}$, following the developments in Eqs. (\ref{eqn:H1})-(\ref{eqn:lambda_herm4}) lead to the simplified formula:
\begin{eqnarray}
 \langle \psi_{i,\mathbf{k}}  \vert p_x \vert \psi_{a,\mathbf{k}}  \rangle \langle \psi_{a,\mathbf{k}}  \vert \frac{\hat{\Omega}_z + \hat{\Omega}^\dagger_z}{2} p_y \vert \psi_{i,\mathbf{k}}  \rangle \nonumber \\
- \langle \psi_{i,\mathbf{k}}  \vert p_y \vert \psi_{a,\mathbf{k}}  \rangle \langle \psi_{a,\mathbf{k}}  \vert \frac{\hat{\Omega}_z + \hat{\Omega}^\dagger_z}{2} p_x \vert \psi_{i,\mathbf{k}}  \rangle
\end{eqnarray} 
where we have introduced the shorthand notation:
\begin{equation}
\hat{\boldsymbol{\Omega}} = \mathbf{r} + \imath \boldsymbol{\nabla_k}
\end{equation}
Repeating the procedure for arbitrary orientations $u$ of the light beam provides:
\begin{eqnarray}
\label{eqn:rotpow}
\left[ \langle \psi_{i,\mathbf{k}}  \vert \mathbf{p} \vert \psi_{a,\mathbf{k}}  \rangle \wedge \langle \psi_{a,\mathbf{k}}  \vert \frac{\hat{\boldsymbol{\Omega}} + \hat{\boldsymbol{\Omega}}^\dagger}{2} \vee \mathbf{p} \vert \psi_{i,\mathbf{k}} \rangle \right]_u \nonumber \\ 
\equiv \sum_{v,w} \epsilon_{u,v,w} \langle \psi_{i,\mathbf{k}}  \vert p_v \vert \psi_{a,\mathbf{k}}  \rangle \langle \psi_{a,\mathbf{k}}  \vert \frac{\hat{\Omega}_u + \hat{\Omega}^\dagger_u}{2} p_w \vert \psi_{i,\mathbf{k}}  \rangle
\end{eqnarray}
for the rotatory strengths, where $\epsilon_{u,v,w}$ is the Levi-Civita symbol and we have introduced a compact notation for the product $\vee$. Eq. (\ref{eqn:rotpow}) includes both DD and DQ contributions, as can be seen by adding and subtracting appropriate matrix elements including $p_u  \frac{\hat{\Omega}_w + \hat{\Omega}^\dagger_w}{2}$:
\begin{eqnarray}
\label{eqn:rotpow2}
\text{DQ} + \text{DD} = \left[ \langle \psi_{i,\mathbf{k}}  \vert \mathbf{p} \vert \psi_{a,\mathbf{k}}  \rangle \wedge \langle \psi_{a,\mathbf{k}}  \vert \frac{\hat{\boldsymbol{\Omega}} + \hat{\boldsymbol{\Omega}}^\dagger}{2} \vee \mathbf{p} \vert \psi_{i,\mathbf{k}} \rangle \right]_u \nonumber \\
= \frac{1}{2}  \sum_{v,w} \epsilon_{u,v,w} \langle \psi_{i,\mathbf{k}}  \vert p_v \vert \psi_{a,\mathbf{k}}  \rangle \nonumber \\
\Big[ \langle \psi_{a,\mathbf{k}}  \vert \frac{\hat{\Omega}_u + \hat{\Omega}^\dagger_u}{2} p_w + p_u  \frac{\hat{\Omega}_w + \hat{\Omega}^\dagger_w}{2} \vert \psi_{i,\mathbf{k}}  \rangle \nonumber \\ 
+ \langle \psi_{a,\mathbf{k}}  \vert \frac{\hat{\Omega}_u + \hat{\Omega}^\dagger_u}{2} p_w  -  p_u  \frac{\hat{\Omega}_w + \hat{\Omega}^\dagger_w}{2} \vert \psi_{i,\mathbf{k}}  \rangle  \Big]
\end{eqnarray}
With the DQ term (i.e. the first term in square brackets) being traceless, orientational averaging (summing over $u=x,y,z$) leads to an expression involving only the DD contribution:
\begin{equation}
\label{eqn:rotpow3}
\langle \psi_{i,\mathbf{k}}  \vert \mathbf{p} \vert \psi_{a,\mathbf{k}}  \rangle \cdot \langle \psi_{a,\mathbf{k}}  \vert \frac{\hat{\boldsymbol{\Lambda}} + \hat{\boldsymbol{\Lambda}}^\dagger}{2} \vert \psi_{i,\mathbf{k}}  \rangle
\end{equation}
We note that Eqs. (\ref{eqn:rotpow})-(\ref{eqn:rotpow3}) contain no terms from the action of $\boldsymbol{\nabla_k}$ on the first-order electric field matrix elements $\langle \psi_{i,\mathbf{k}} \vert \mathbf{p} \vert \psi_{a,\mathbf{k}}  \rangle$. Just like for the magnetization from Eq. (\ref{eqn:m_gauge}), the rotatory strengths become independent of the gauge-origin and orbital phase at the CBS limit. This may be seen by inserting Eq. (\ref{eqn:rep_identity}), as well as Eq. (\ref{eqn:rep_nabla}) into Eq. (\ref{eqn:rotpow2}), yielding:
\begin{eqnarray}
 \text{DQ} + \text{DD}  &=& \sum_l^\text{all} \Big[ - \langle \psi_{i,\mathbf{k}}  \vert \mathbf{p} \vert \psi_{a,\mathbf{k}}  \rangle \wedge \frac{\langle \psi_{a,\mathbf{k}}  \vert \boldsymbol{\nabla_r}  \vert \psi_{l,\mathbf{k}} \rangle}{ \epsilon_{i,\mathbf{k}} - \epsilon_{l,\mathbf{k}}  } \nonumber \\
&\vee& \langle \psi_{l,\mathbf{k}}  \vert  \mathbf{p} \vert \psi_{i,\mathbf{k}} \rangle \Big]_u + \text{H.c.}
\end{eqnarray}
which is periodic, and does not depend on the gauge origin, or the gradient of the orbital phases $- \boldsymbol{\nabla_k} \phi_i \left( \mathbf{k} \right)$.

Finally, in recent work on OR, Wang and Yan find a new ``band dispersion'' contribution to the rotatory strengths. In Appendix \ref{app:band_disp} we show how this band dispersion term is an alternate formulation of the magnetic dipole term in Eq. (\ref{eqn:rotpow2})for semi-local functionals. As a consequence, band dispersion contributions are not included in our formulation.

\subsection{Optical Rotation Tensor}

In the electronic supporting information (ESI) we develop a formula for the diagonal elements of the OR tensor $\beta_{u}$ (including both magnetic-dipole as well as electric-quadrupole contributions) based on the operators of Eq. (\ref{eqn:rotpow}) and time-dependent double-perturbation theory. 

$\boldsymbol{\beta}$ relates the components of the magnetic-dipole plus electric-quadrupole moments induced by the electric field $\boldsymbol{\eta}$, to the time-derivative of the electric field $\boldsymbol{\mathcal{E}}$ (see ESI for more details):\cite{condon1937theories}
\begin{equation}
\boldsymbol{\eta} = \frac{\boldsymbol{\beta}}{c} \frac{\partial}{\partial t} \boldsymbol{\mathcal{E}} \left( t,\omega \right)
\end{equation}
For practical calculations, we expand the COs in a finite set of functions $\varphi_{\mu,\mathbf{k}}$ labelled by the index $\mu$:
\begin{equation}
\label{eqn:lcao}
\vert \psi_{l,\mathbf{k}} \rangle = \sum_\mu C_{\mu,l} \left( \mathbf{k} \right) \vert \varphi_{\mu,\mathbf{k}} \rangle
\end{equation}
with the CO coefficients $C_{\mu,l} \left( \mathbf{k} \right)$ being determined from the solution of the field-free GKS-DFT equations. To write a compact expression for $\beta_{u}$, it is convenient to use an analytical expression for the derivatives of the coefficients $C_{\mu,l} \left( \mathbf{k} \right)$ in terms of a matrix $\mathbf{Q}$:\cite{otto1999calculation}
\begin{equation}
\label{eqn:Qmat}
\boldsymbol{\nabla_k} C_{\mu,l} \left( \mathbf{k} \right) = \sum_{l^\prime}^\text{all} \mathbf{Q}_{l^\prime,l} \left( \mathbf{k} \right)  C_{\mu,l^\prime} \left( \mathbf{k} \right) 
\end{equation}
Similarly, the $u$-component electric-field perturbed coefficients $C_{\mu,i}^{(u),\pm} \left( \mathbf{k} \right)$, are written in terms of a matrix $\mathbf{U}$ (in the non-canonical treatment only anti-Hermitian virt-occ interbank elements $U_{ai}^{(u),\pm}$ are non-vanishing):\cite{karna1991frequency}
\begin{equation}
\label{eqn:Umat}
C_{\mu,i}^{(u),\pm} \left( \mathbf{k} \right) = \sum_{a}^\text{virt} U_{ai}^{(u),\pm} \left( \mathbf{k} \right) C_{\mu,a} \left( \mathbf{k} \right) 
\end{equation}
In Eq. (\ref{eqn:Qmat}) $\mathbf{Q}$ is determined from the derivative of the GKS equation and orthonormality condition:\cite{otto1999calculation}
\begin{subequations}
\begin{equation}
\label{eqn:Q}
 \mathbf{Q}_{l,l^\prime} (\mathbf{k}) = \frac{ \mathbf{K}_{l,l^\prime} \left( \mathbf{k} \right) - \epsilon_{l ^\prime,\mathbf{k}} \mathbf{R}_{l,l^\prime} \left( \mathbf{k} \right) }{ \epsilon_{l^\prime,\mathbf{k}} - \epsilon_{l,\mathbf{k}}} 
\end{equation}
for off-diagonal $l \ne l^\prime$ elements and:
\begin{equation}
\Re \left[ \mathbf{Q}_{l,l} (\mathbf{k}) \right] = - \frac{1}{2} \mathbf{R}_{l,l} \left( \mathbf{k} \right) 
\end{equation}
\end{subequations}
for diagonal elements. Here $\mathbf{K}_{l,l^\prime} \left( \mathbf{k} \right)$ and $\mathbf{R}_{l,l^\prime} \left( \mathbf{k} \right)$ are the $\mathbf{k}$-derivatives of the GKS-DFT Hamiltonian and overlap matrices, respectively, for fixed CO coefficients. The imaginary part of $\mathbf{Q}_{l,l} (\mathbf{k})$, on the other hand, is determined by integers associated with the phase function $\phi \left( \mathbf{k} \right)$ of Eq. (\ref{eqn:CO_BF}),\cite{bishop2001coupled} which we set to zero, as discussed in section \ref{sec:gauge}. 

In our implementation, we work with the Hermitian quantity:
\begin{equation}
\label{eqn:tildeQ}
\tilde{\mathbf{Q}}_{l,l^\prime} (\mathbf{k}) = \mathbf{Q}_{l,l^\prime} (\mathbf{k}) + \frac{1}{2} \mathbf{R}_{l,l^\prime} \left( \mathbf{k} \right)
\end{equation}
In terms of these coefficients, the final formula for $\beta_{u}$ reported in Eq. (S45) is given by:
\begin{eqnarray}
\label{eqn:beta1}
\beta_{u} &=& \frac{1}{\omega} \text{P.V.} \int^\prime d \mathbf{k} \sum_i^\text{occ} \sum_a^\text{virt} \Im \Bigg[  \left( \mathbf{U}_{ai}^{+} \left( \mathbf{k} \right) - \mathbf{U}_{ai}^{-} \left( \mathbf{k} \right)  \right)  \nonumber \\
&\wedge& \left( \tilde{\mathbf{v}}_{ia} \left( \mathbf{k} \right) + \mathbf{q}_{ia} \left( \mathbf{k} \right) \right)  \Bigg]_u
\end{eqnarray}
where
\begin{eqnarray}
\mathbf{q}_{ia} \left( \mathbf{k} \right) &=& \frac{1}{2} \sum_l^\text{all} \Big[ \left( \tilde{\mathbf{Q}}_{l,i} (\mathbf{k}) - \frac{1}{2} \mathbf{R}_{l,i} \left( \mathbf{k} \right) \right) \vee \langle \psi_{a,\mathbf{k}} \vert \boldsymbol{\nabla}_{r} \vert \psi_{l,\mathbf{k}} \rangle \nonumber \\
&-& \left( \tilde{\mathbf{Q}}_{l,a}^\ast (\mathbf{k}) - \frac{1}{2} \mathbf{R}_{l,a}^\ast \left( \mathbf{k} \right)  \right) \vee \langle \psi_{l,\mathbf{k}} \vert \boldsymbol{\nabla}_{r} \vert \psi_{i,\mathbf{k}} \rangle \Big]
\end{eqnarray}
and $\mathbf{v}_{ia}$ is defined below in Eq. (\ref{eqn:via}).

In Eq. (\ref{eqn:beta1}), the prime indicates that integration is restricted to the portion of the FBZ with positive coordinates (we have made use of the fact that the operators of Eq.(\ref{eqn:rotpow}) are odd under inversion $\mathbf{k} \to -\mathbf{k}$) and  $\text{P.V.}$ indicates that the integral must be interpreted in terms of its Cauchy principal value (or, equivalently, its finite part). Thus, the elements of $\tilde{\mathbf{Q}}$ are calculated as:
\begin{equation}
\label{eqn:Qeta}
\tilde{\mathbf{Q}}_{l,l^\prime} (\mathbf{k}) = \lim_{\eta \to 0^+} \frac{ \mathbf{K}_{l,l^\prime} \left( \mathbf{k} \right) - \frac{1}{2} \left( \epsilon_{l,\mathbf{k}} + \epsilon_{l ^\prime,\mathbf{k}} \right)\mathbf{R}_{l,l^\prime} \left( \mathbf{k} \right) }{ \epsilon_{l^\prime,\mathbf{k}} - \epsilon_{l,\mathbf{k}} +\imath \eta }
\end{equation}
for $l \ne l^\prime$ with $\eta$ being a small positive number. The parameter $\eta$ provides an effective means to deal with quasi-degeneracies for occ-occ or virt-virt blocks of $\tilde{\mathbf{Q}}$ wherein the denominator in Eq. (\ref{eqn:Qeta}) vanishes in the limit $\eta \to 0$. In section \ref{sec:res} we demonstrate that accurate calculations are not sensitive to the value of $\eta$, provided that it is sufficiently small. Finally, in Eq. (\ref{eqn:beta1}) $\tilde{\mathbf{v}}_{ia}=\frac{\mathbf{v}_{ia}+\mathbf{v}_{ai}^\ast}{2}$, where $\mathbf{v}_{ia}$ reads:
\begin{equation}
\label{eqn:via}
\mathbf{v}_{ia} = \langle \psi_{i,\mathbf{k}}  \vert \left( \mathbf{r} + \mathbf{R}_O \right) \vee \mathbf{p} \vert \psi_{a,\mathbf{k}} \rangle
\end{equation}
and $\mathbf{R}_O$ is the gauge origin.

The $U_{ai}^{(u),\pm} \left( \mathbf{k} \right)$ matrix elements of Eq. (\ref{eqn:Umat}) for the $u$-component of the electric-dipole moment, that are used in sum-over-states (SOS) calculations, are:
\begin{equation}
\label{eqn:u_sos}
U_{ai}^{(u),\pm} \left( \mathbf{k} \right) = \frac{ \langle  \psi_{a,\mathbf{k}} \vert \Omega_u \vert \psi_{i,\mathbf{k}} \rangle }{\epsilon_{a,\mathbf{k}} - \epsilon_{i,\mathbf{k}} \pm \omega} \simeq -\frac{ \langle  \psi_{a,\mathbf{k}} \vert \nabla_u \vert \psi_{i,\mathbf{k}} \rangle }{\left( \epsilon_{a,\mathbf{k}} - \epsilon_{i,\mathbf{k}} \pm \omega \right)\left(  \epsilon_{a,\mathbf{k}} - \epsilon_{i,\mathbf{k}} \right)}
\end{equation}
For TDDFT calculations these must be augmented with orbital-relaxation contributions, calculated here through a self-consistent solution of the coupled-pertubed generalized Kohn-Sham (CPKS) equations.\cite{rerat2021first,ferrari2015ab} In Eq. (\ref{eqn:u_sos}), the last equality implies validity of the off-diagonal hypervirial relation, which holds only for calculations employing semi-local Hamiltonians (or a complete basis). We denote calculations with the operator $\Omega_u$on the left of the last equality as the length (or L) formulation, and those with the $\nabla_u$ operator as the velocity (or V) formulation.

As discussed in R{\'e}rat and Kirtman,\cite{rerat2021first} the OR tensor is independent of $\mathbf{R}_O$ by construction if the velocity operator is used in Eq. (\ref{eqn:u_sos}). In that case we set $\mathbf{R}_O=\mathbf{0}$ (V0 formulation). On the other hand, calculations with the length operator are generally gauge-origin dependent, and, then, we use the electronic centroid $\mathbf{R}_O=\int_\text{BZ} d \mathbf{k} \sum_i^\text{occ} \langle \psi_{i,\mathbf{k}} \vert \mathbf{r} \vert \psi_{i,\mathbf{k}} \rangle$ as the gauge-origin (LC formulation).

\section{Computational Details}
\label{sec:comp}
Unless explicitly stated otherwise, all calculations were performed with a developer's version of the \textsc{Crystal23} code,\cite{erba2022crystal23} employing all-electron Gaussian atomic-orbital (AO) basis sets. The field-free calculations were converged down to a criterium of 1$\times$10$^{-10}$ Hartree a.u. ($E_h$) on the total energy.

The xc contribution was calculated by numerical quadrature using Gauss-Legendre radial and Lebedev angular point distributions,\cite{towler1996density,lebedev1976quadratures,lebedev1977spherical} with the quadrature weights proposed by Becke.\cite{becke1988multicenter} We used a pruned grid consisting of 99 radial points and 1454 angular points (keyword \textsc{XXLGRID} in the \textsc{Crystal23} manual).\cite{crystal23_man} The \textsc{XcFun} library\cite{ekstrom2010arbitrary} was employed for taking the SVWN5 (local density approximation, LDA), PBE (generalized gradient approximation, GGA) and PBE0 (hybrid approximation) xc functional derivatives required for the first-order CPKS procedure.\cite{s_art,vwn_art,pbe_art,pbe0_art} More specific details are available from the ESI, where the full input decks are provided.

\section{Results and Discussion}
\label{sec:res}

The implementation is validated in several respects on chains of H$_2$O$_2$, which has served as a model system in previous work.\cite{balduf2022derivation,rerat2021first} In detail, we discuss:

\begin{enumerate}
\item the effect of the parameter $\eta$ of Eq. (\ref{eqn:Qeta}) for dealing with quasi-degeneracies in the calculation of $\mathbf{k}$-derivatives of the unperturbed orbital coefficients
\item comparisons of infinite against finite chains, where matching results against the large finite system is obtained as the CBS limit is approached
\item the equivalence of sampling reciprocal space (number of $\mathbf{k}$ points in the FBZ) and direct space (size of the supercell) being verified as the CBS limit is approached
\end{enumerate}

Points 2. and 3. were anticipated from the discussion in section \ref{sec:gauge}. After validation of our approach, we present applications to the calculation of OR from linear-response (LR) TDDFT in the adiabatic approximation with hybrid functionals in $\alpha$-quartz, where the importance of i) DQ terms, ii) orbital relaxation, iii) non-local Fock exchange, and iv) completeness of the basis set expansion is discussed.

\subsection{Validation on Finite and Infinite Chains of H$_2$O$_2$}

The calculations on H$_2$O$_2$ chains were performed with geometries reported in Ref. \onlinecite{rerat2021first} and available from the input decks in the ESI. The correlation-consistent polarized valence family of basis sets of Dunning were employed, with double, triple and quadruple zeta (cc-pvXz, with X = D, T, Q).\cite{dunning1989gaussian} We do not report on calculations beyond quadruple zeta, because we obtained quasi-linear dependencies with quintuple zeta and larger basis sets, and corresponding results, then, depended heavily on the overlap eigenvalue threshold for canonical orthonormalization. The comparison of finite vs. infinite H$_2$O$_2$ chains was done with the SVWN5 functional of the LDA, and SOS-V0 formulation, being the computationally simplest case that allows us to discuss all relevant aspects. We report calculations of the mean OR, which is calculated from the DD, but not DQ term, as per Eq. (\ref{eqn:rotpow3}), so as to discuss calculation of matrix-elements of the Hermitized angular momentum $\hat{\boldsymbol{\Lambda}}+\hat{\boldsymbol{\Lambda}}^\dagger$ operator.

\begin{figure}[t!]
\centering
\includegraphics[width=8.6cm]{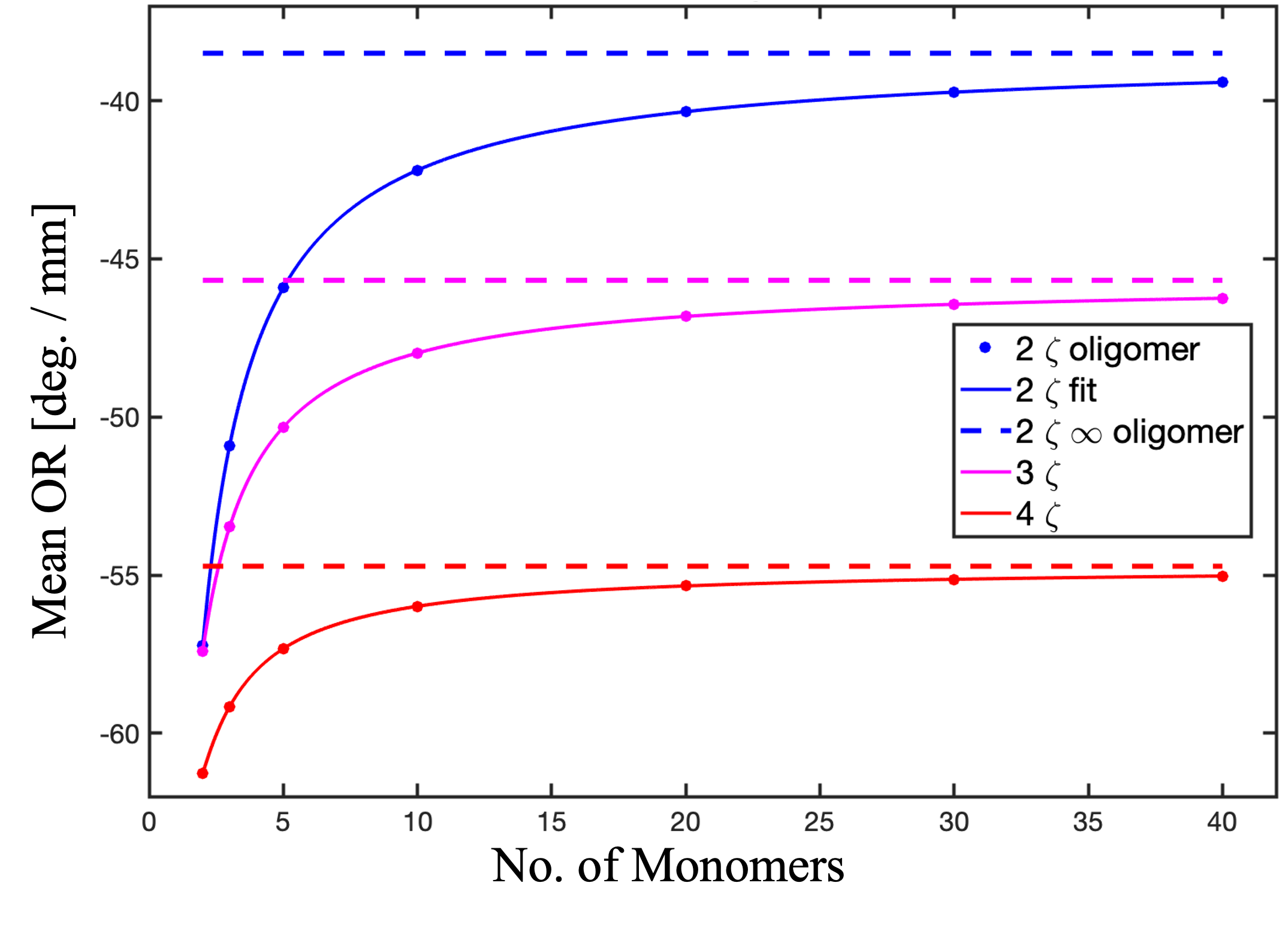}
\caption{Mean OR of finite H$_2$O$_2$ oligomers (dots), employing the cc-pvXz (X = D, T, Q) basis sets (blue, magenta, red), polynomial fits (solid lines) and values extrapolated to infinite chain length (dashed lines).}
\label{fig:mean_h2o2}
\end{figure}

Fig. \ref{fig:mean_h2o2} provides values of the mean OR of H$_2$O$_2$ chains, as a function of chain length and size of the single-particle basis expansion. The solid lines represent fits to polynomials of the form $c_4/x^4+c_3/x^3+c_2/x^2+c_1/x+c_0$. The value of the fitted $c_0$ coefficients (representing the reference finite oligomer value, extrapolated to infinite chain length) are plotted in the dashed lines, being $38.50, 45.67$ and $54.72 \ ^\circ / mm$ for X = D, T, Q. 

To confirm the quality of the single-particle basis set expansion, in Table \ref{fig:ao_giao} we provide length and velocity results on finite oligomers with AO, as well as gauge-including AO (GIAO) basis sets. The GIAO results were obtained with the implementation in the Gaussian program.\cite{g16} Of course, all values for a given system should coincide at the CBS limit. The good agreement with the cc-pvQz basis set (last row of the table) is indicative that the calculations are nearly converged to the CBS limit. Although length and velocity gauge cc-pvQz results still show considerable disparity on the isolated H$_2$O$_2$ molecule (66.83 vs. 54.66), the differences become much smaller for longer chain lengths (e.g. 55.35 vs. 55.24 for the 20-unit oligomer).

\begin{table}[t]
\scriptsize
\caption{Mean OR as calculated with the V0 and LC formulations and an atomic orbital (AO) and gauge-including atomic orbital (GIAO) basis on the isolated H$_2$O$_2$ molecule and 10-unit and 20-unit oligomer. Calculations on the isolated molecule included orbital relaxation to allow comparison against the GIAO implementation.}
\vspace{5pt}
\begin{tabular}{ll | ccc c cc c cc}
\hline
\hline
&&\\
& & \multicolumn{3}{c}{1$\times$H$_2$O$_2$} && \multicolumn{2}{c}{10$\times$H$_2$O$_2$} && \multicolumn{2}{c}{20$\times$H$_2$O$_2$}\\
& cc-pvXz & LC-AO & V0-AO & L-GIAO  && LC-AO & V0-AO && LC-AO & V0-AO \\
\hline
&&\\
& D  & 70.45 & 75.61 & 59.24  && 21.56 & 42.20 && 27.56 & 40.34 \\
& T  & 65.99 & 59.67 & 54.53  && 32.08 & 47.98 && 29.43 & 46.82\\
& Q  & 66.83 & 54.66 & 55.32  && 56.67 & 55.99 && 55.35 & 55.24\\
&&\\
\hline
\hline
\label{fig:ao_giao}
\end{tabular}
\end{table}
\normalsize

Next, we turn to periodic calculations on the infinite H$_2$O$_2$ chain. As per, Eqs. (\ref{eqn:beta1})-(\ref{eqn:Qeta}), the periodic formulation requires a careful treatment of quasi-degeneracies in the calculation of $\boldsymbol{\nabla_k} C_{\mu,l} \left( \mathbf{k} \right)$. Of course, for an exact solution of the GKS-DFT equations the $C_{\mu,l} \left( \mathbf{k} \right)$ are differentiable,\cite{blount1962formalisms} and, then, $\tilde{\mathbf{Q}}$ in Eq. (\ref{eqn:Qeta}) must be finite, even at quasi-degeneracies. Practical calculations, on the other hand, are approximate solutions of the GKS-DFT equations, and, then, require a judicious choice of the $\eta$ parameter of Eq. (\ref{eqn:Qeta}) for systems that contain quasi-degeneracies in the occupied and/or virtual band structure. As a test for systems with quasi-degeneracies, we construct supercells of the H$_2$O$_2$ polymer (the supercell introduces quasi-degneracies in the FBZ) and report on values of the mean OR for $\times$7, $\times$9 and $\times$11 supercell $\Gamma$-point calculations in Table \ref{tab:eta}. The table confirms that the mean OR is not sensitive to the precise value of  $\eta$, so long as it is sufficiently small (at least $10^{-14}$ $E_h$ a.u. in this case). Based on the results of Table \ref{tab:eta}, we employ the value $\eta = 10^{-14}$ $E_h$ a.u. in all subsequent calculations. With this value of $\eta$, the $\Gamma$-point results match the infinite oligomer value of $38.50$ $^\circ / mm$ for the sufficiently large $\times$ 9 and $\times$ 11 supercells. Although this choice of $\eta$ allows us to achieve the necessary level of numerical accuracy, convergence of the calculated values w.r.t., e.g. number of $\mathbf{k}$-points and other numerical tolerances is relatively slow. To improve the implementation, a more elaborate scheme for a treatment of quasi-degeneracies in the calculations of derivatives of the CO coefficients $\boldsymbol{\nabla_k} C_{\mu,l} \left( \mathbf{k} \right)$ will be presented in a forthcoming publication, in conjunction with new applications.

\begin{table}[t]
\caption{$\Gamma$-point calculation of the mean OR (in $^\circ / mm$) on supercells of an H$_2$O$_2$ polymer with the cc-pvDz basis set. The $\infty$-oligomer reference value is $38.499 \ ^\circ / mm$}
\vspace{5pt}
\begin{tabular}{llcccccccccc}
\hline
\hline
&&\\
& $\eta$ & $10^{-8}$ & $10^{-10}$ & $10^{-12}$ & $10^{-14}$ & $10^{-16}$ & $10^{-75}$ & $10^{-150}$ \\
\hline
&&\\
& $\times$ 7  & 49.986 & 44.537 & 38.654 & 38.506 & 38.506 & 38.506& 38.506 \\
& $\times$ 9  & 50.584 & 45.504 & 38.502 & 38.499 & 38.499 & 38.499& 38.499 \\
& $\times$ 11 & 50.530 & 44.557 & 35.315 & 38.499 & 38.499 & 38.499& 38.499 \\
&&\\
\hline
\hline
\label{tab:eta}
\end{tabular}
\end{table}

With the choice of $\eta$ being established, we perform calculations on the small cell of H$_2$O$_2$ with many $\mathbf{k}$-points and compare sampling of reciprocal (number of $\mathbf{k}$-points in the FBZ) and direct (size of the supercell expansion) spaces in Table \ref{tab:gk}. For the small cell, results are fully converged with 500 $\mathbf{k}$ points. 

Table \ref{tab:gk} confirms the analysis of section \ref{sec:gauge}. Generally, in a finite basis $\hat{\boldsymbol{\Lambda}}+\hat{\boldsymbol{\Lambda}}^\dagger$ is not periodic and only becomes periodic at the CBS limit. As a consequence, significant differences between the small cell vs. infinite oligomer results (first and last columns of the table, i.e. 47.90 vs. 38.50 $^\circ / mm$) are obtained with the double-zeta basis set. The difference is greatly improved at the triple-zeta level to 47.88 vs. 45.67 $^\circ / mm$. Finally, essentially matching results (53.55 vs. 54.72 $^\circ / mm$) are gotten with the quadruple-zeta basis set.

Again, because of the non-periodicity of $\hat{\boldsymbol{\Lambda}}+\hat{\boldsymbol{\Lambda}}^\dagger$, disparities can be observed for sampling of reciprocal vs. direct space, especially with a small basis set. Concentrating on the 3/3 vs. 9/1 calculations (second and fourth columns of the table) significant differences (44.07 vs. 38.50 $^\circ / mm$) are obtained with a double-zeta basis set, which is improved to 46.14 vs. 45.67 $^\circ / mm$ at the triple-zeta level, and finally almost matching results of 54.25 vs. 54.72 $^\circ / mm$ are obtained at the quadruple-zeta level. 
As mentioned in section \ref{sec:gauge}, the non-periodicity of $\hat{\boldsymbol{\Lambda}}+\hat{\boldsymbol{\Lambda}}^\dagger$  is associated with its gauge-origin dependence. In the case of finite systems, the problem may be solved by including field-dependent phase factors in the basis functions (GIAOs).\cite{mcweeny1971gauge,peters2022magnetic} Whether a related, or alternative, approach to the problem can be developed for infinite, periodic systems remains to be demonstrated.

\begin{table}[t]
\caption{Mean OR in $[^\circ / mm]$ for different supercells, number of $\mathbf{k}$ points and basis sets, as compared to the reference infinite oligomer values.}
\vspace{5pt}
\begin{tabular}{llccccccc}
\hline
\hline
&&\\
& cells/$\mathbf{k}$-points & 1/500 & 3/3 & 7/1 & 9/1 & 11/1 && $\infty$-oligo. \\
\hline
&&\\
& cc-pvDz  & 47.90 & 44.07 & 38.51 & 38.50 & 38.50 && 38.50\\
& cc-pvTz  & 47.88 & 46.14 & 45.68 & 45.67 & 45.67 && 45.67\\
& cc-pvQz  & 53.55 & 54.25 & 54.75 & 54.72 & 54.72 && 54.72\\
&&\\
\hline
\hline
\label{tab:gk}
\end{tabular}
\end{table}

\subsection{Application to $\alpha$-quartz}

Quartz is the dominant mineral composing the Earth's outer crust. The low-temperature $\alpha$-quartz phase cristallizes in a trigonal space group, in either right- or left-handed helix (space groups $P 3_1 21$ and $P 3_2 21$) polymorphs, which may be distinguished experimentally on the basis of their optical rotatory power.\cite{skalwold2015quartz} 

Here, we report first-principles calculations of the OR in right $\alpha$-quartz (i.e. $P 3_2 21$ dextrorotatory $\alpha$-quartz with clockwise rotation of the plane of polarization when facing the source of light). Indeed, this sign convention for the OR angle about direction $u$, provides:\cite{buckingham1971optical}
\begin{equation}
\label{eqn:phi_n}
\Phi_u = \frac{(n_L - n_R) \pi}{\lambda}
\end{equation}
where $n_L$ ($n_R$) is the refractive index for left (right) circularly-polarized light in the ($v,w$)-plane. Then, $\Phi_u$ is negative if $n_R > n_L$, i.e. the rotation of the plane of polarization is towards the right (dextro-rotatory) for light coming towards the observer.

We obtain converged results with a dense $24 \times 24 \times 24$ Monkhorst-Pack net for sampling of reciprocal space (see ESI for full input decks). To assess the effect of the quality of the single-particle basis set, we performed calculations with the double and triple-zeta valence polarization basis sets of Peintinger, Oliveira and Bredow (POB-DZVP and POB-TZVP),\cite{peintinger2013consistent} as well as the 6-311G(d) basis set of Heyd and co-workers,\cite{heyd2005energy} and quasi-relativistic effective-core potentials and valence basis sets, from adjustment to the Wood-Boring Hamiltonian (ECP-MWB).\cite{bergner1993ab} Crystal structures were fully optimized, under constraints provided by the trigonal space group, with the PBE0 functional. The optimized lattice parameters and GKS indirect band gaps are reported in Table \ref{tab:quartz_struct}, as compared with values from high-resolution powder X-ray diffraction and electron energy loss spectroscopy experiments and photoconductivity measurements.\cite{antao2008state,garvie2000bonding,distefano1971band,evrard1982photoelectric} Calculations with all four basis sets provide GKS gaps (7.639 to 9.220 eV) that are somewhat underestimated, compared to the experimental value of 8.9 to 11.5.\cite{antao2008state,garvie2000bonding,distefano1971band,evrard1982photoelectric} This may be expected, as the calculated gaps from GKS eigenvalue differences are largely dependent on the fraction of Fock exchange and neglect exciton effects.\cite{ferrari2015ab} For the lattice parameters, a good agreement against the experiment is obtained with the Heyd basis sets. It is also noteworthy that the total energy is lowest with the Heyd basis set. 

\begin{table}[t]
\caption{Optimized lattice parameters [Angstrom], indirect band gaps [eV] of $\alpha$-quartz with the PBE0 functional and different basis sets, as compared to experimental values.\cite{antao2008state,garvie2000bonding,distefano1971band,evrard1982photoelectric} Total energy differences $\Delta E_\text{Heyd}$ are also reported w.r.t. the Heyd 6-311G(d) value.}
\vspace{5pt}
\begin{tabular}{ll | cccc}
\hline
\hline
&&\\
&& a & c & gap & $\Delta E_\text{Heyd}$ \\
\hline
&&\\
& ECP-MWB     & 5.209 & 5.735 & 7.639 & -    \\
& POB-DZVP    & 5.019 & 5.546 & 9.220 & 0.749\\
& POB-TZVP    & 4.998 & 5.481 & 8.837  & 0.011\\
& Heyd 6-311G(d)  & 4.927 &  5.431 & 8.431 & 0.000\\
& Experiment & 4.913 & 5.405 & 8.9-11.5 & - \\
&&\\
\hline
\hline
\label{tab:quartz_struct}
\end{tabular}
\end{table}

At the previously optimized PBE0 geometries, we then calculated the OR tensor, employing the SOS and LR-TDDFT approaches (i.e. without and with account of orbital relaxation contributions), using the PBE and PBE0 functionals. Since $\alpha$-quartz, is a uniaxial positive mineral, the optic axis lies parallel to the $c$ crystallographic axis and experimental measurements are available at a wavelength of $\lambda = 589.44 \ nm$, yielding a value $\Phi_c=-21.7 \ ^\circ / mm$.\cite{skalwold2015quartz} 

Calculated values with the length (LC) and velocity (V0) gauges for the electric dipole operator are reported in Table \ref{tab:quartz}. The table shows that both an inclusion of Fock exchange and orbital relaxation are crucial in the calculation of the OR tensor. As far as orbital relaxation is concerned, a comparison of SOS vs. LR-TDDFT values shows a large difference  -- e.g. +9.01 to -2.23 for PBE and +3.49 to +0.10 for PBE0 with V0 and the DZVP basis set. The most extreme case occurs with the Heyd basis set, orbital relaxation results in a change of $\Phi_c$ from  +4.39 to -63.22  for PBE and +0.97 to -26.14 for PBE0. The effect of including 25 \% Fock exchange in the functional is also quite important; in the V0 case, the LR-TDDFT values go from -2.23 to +0.10 for DZVP, -2.31 to -0.042 for TZVP, and -63.22 to -26.14 for the Heyd G-311G(d) basis. 

One way of assessing the quality of the single-particle basis is by comparing relative LR-TDDFT values in the length and velocity gauges. In principle, the LC and V0 values should match at the CBS limit. A small variation, therefore, indicates a good quality basis. The relative variation is much smaller for the Heyd basis set than for the ECP-MWB, POB-DZVP or POB-TZVP ones. This result agrees with the preceding calculations of total energies and lattice parameters, but not the indirect gaps. 


For our best calculation (PBE0 functional, Heyd basis set, LR-TDDFT treatment, gauge-origin invariant V0 formulation), we obtain $\Phi_c=-26.14 \ ^\circ / mm$, in good agreement with the experimental value of $\Phi_c=-21.7 \ ^\circ / mm$. For this particular calculation, the individual contributions from DD and DQ terms to the total $\Phi_c$ are $\text{DD} = -14.62 \ ^\circ / mm$ and $\text{DQ} = -11.52 \ ^\circ / mm$, thereby confirming the importance of DQ terms for calculations in the solid state, as also stressed in two recent other studies.\cite{balduf2022derivation,wang2022optical} 

Although the Heyd basis set yields reasonably good agreement with experiment, Table \ref{tab:quartz} displays a large dependence on the gauge (LC vs. V0), as well as the basis set. To improve this dependence, in the future we will consider the possibility of modifying the single-particle orbitals by field-dependent phase factors. Indeed, the OR values are very sensitive to the presence and exact nature of diffuse functions in the basis set. Removal of the most diffuse $s$ and $p$ functions in the Heyd basis set changes the PBE0 LR-TDDFT values obtained with the V0 (LC) formulation from $\Phi_c=$-26.14 (-36.05) $\ ^\circ / mm$ to $\Phi_c=$+252.02 (-4.17) $\ ^\circ / mm$. Finally, preliminary analysis indicates that the effect of the basis set is influenced in a major way by its capacity to yield accurate geometries. For instance, the PBE0 LR-TDDFT values obtained with the POB-TZVP basis set of +3.37 (-0.042) $^\circ / mm$ with the V0 (LC) formulation are much improved to -4.98 (-5.01) $^\circ / mm$  by using the optimized geometry from the Heyd basis set. The same values are further improved to -8.54 (-8.84) $^\circ / mm$ by addition of a $p$ function with exponent of 0.12 \text{bohr}$^{-2}$ on the O atoms. Good agreement against the experiment on OR, therefore requires good agreement on optimized geometries.

\begin{table}[t]
\caption{PBE and PBE0 calculation of $\Phi_c$  (in $^\circ / mm$) on $\alpha$-quartz at wavelength $\lambda = 589.44 \ nm$ with the SOS and LR-TDDFT approaches (i.e. without and with account of orbital relaxation contributions) for different basis sets. The experimental value is $\Phi_c=-21.7 \ ^\circ / mm$.\cite{skalwold2015quartz}}
\vspace{5pt}
\begin{tabular}{ll |   cc c cc}
\hline
\hline
&&\\
&& \multicolumn{5}{c}{PBE} \\
&& \multicolumn{2}{c}{SOS} && \multicolumn{2}{c}{LR-TDDFT} \\
& basis set & LC & V0  && LC & V0 \\
\hline
&&\\
& ECP-MWB     & -20.98 & +21.20 && -43.61& -5.53 \\
& POB-DZVP    & +10.35 & +9.01 &&  +3.24 & -2.23 \\
& POB-TZVP    & +7.90 & +6.64 &&  +1.12 &  -2.31  \\
& Heyd 6-311G(d) & -6.96 & +4.39 && -56.29 & -63.22 \\
&&\\
&& \multicolumn{5}{c}{PBE0} \\
&& \multicolumn{2}{c}{SOS} && \multicolumn{2}{c}{LR-TDDFT} \\
& basis set & LC & V0  && LC & V0 \\
\hline
&&\\
& ECP-MWB     & -13.43& +7.73 &&  -31.48& -1.96 \\
& POB-DZVP    & +4.81 & +3.49 &&  +0.80 & +0.10 \\
& POB-TZVP    & +2.42 & +4.27 && +3.37 & -0.042 \\
& Heyd 6-311G(d) & -1.72 & +0.97 &&   -36.05 & -26.14 \\
&&\\
\hline
\hline
\label{tab:quartz}
\end{tabular}
\end{table}

\section{Conclusions}
\label{sec:concu}
A previously proposed electronic ``angular-momentum'' operator is shown to generalize the ``modern theory of orbital magnetization'' to non-local Hamiltonians (e.g. hybrid exchange-correlation functionals of generalized Kohn-Sham theory). A rigorous development of the theory demonstrates that previously suggested  ``band dispersion'' terms, as well as terms involving reciprocal space derivatives of the perturbed wave function, can be avoided. Finally, it is shown that while the ``angular momentum'' operator is (in principle) periodic at the complete basis set limit, it is not so in the general case. 

The formalism is applied to calculating the optical rotatory power (OR) of band insulators in the public \textsc{Crystal} program, where expressions are developed in terms of the electric dipole - electric quadrupole (DQ) as well as the electric dipole - magnetic dipole (DD) contributions. For effective calculations, a strategy is developed to deal with quasi-degeneracies in obtaining  derivatives of the orbital coefficients with respect to the wave-vector. Our implementation is validated by comparison with a model finite system and we report on an application to the $\alpha$-quartz mineral, with linear-response time-dependent density functional theory calculations, that employ a hybrid as well as a non-hybrid functional. This application confirms the importance of DQ terms for OR calculations in the solid state. In the case of $\alpha$-quartz, agreement versus experiment was only possible with an explicit account of i) use of a high quality basis set ii) inclusion of a fraction of non-local exact exchange in the exchange-correlation functional and iii) taking account of orbital-relaxation.

For an implementation that is less dependent on the number of $\mathbf{k}$ points used to sample reciprocal space, as well as other numerical parameters, we will present in the future an improved scheme for the treatment of quasi-degeneracies in determining derivatives of the orbital coefficients with respect to the wave-vector. This will be done in conjunction with new applications for periodic systems of interest.

\appendix
\numberwithin{equation}{section}
\section{Derivation of Eq. (\ref{eqn:u_Hgradk1})}
\label{app:Hgradk}
Expanding the matrix element of Eq. (\ref{eqn:MLC}) in WFs using Eqs. (\ref{eqn:dkbloch_w}) and (\ref{eqn:Hkdkbloch_w}) gives:
\begin{eqnarray}
\label{eqn:MLC_long}
\mathbf{M}_\text{LC} = \frac{1}{2c} \Re \sum_i^\text{occ} \int_{\text{BZ}} d \mathbf{k} \ \sum_{\mathbf{g} \mathbf{g}^\prime} e^{-\imath \mathbf{k} \cdot \left( \mathbf{g}^\prime - \mathbf{g} \right)} \nonumber \\
\times \langle w_{i,\mathbf{g}^\prime}  \vert \left( \mathbf{r} - \mathbf{g}^\prime \right)  \wedge H / \imath \left( \mathbf{r} - \mathbf{g} \right) \vert w_{i,\mathbf{g}}  \rangle \nonumber \\
= \frac{1}{2c \Omega} \Re \sum_i^\text{occ} \sum_{\mathbf{g} \mathbf{g}^\prime} \delta_{\mathbf{g} \mathbf{g}^\prime} \langle w_{i,\mathbf{g}^\prime}  \vert \left( \mathbf{r} - \mathbf{g}^\prime \right)  \wedge H / \imath \left( \mathbf{r} - \mathbf{g} \right) \vert w_{i,\mathbf{g}}  \rangle \nonumber \\
= - \frac{1}{2c \Omega} \Re \sum_i^\text{occ} \sum_{\mathbf{g}} \langle w_{i,\mathbf{g}}  \vert \left( \mathbf{r} - \mathbf{g} \right)  \wedge \imath H \left( \mathbf{r} - \mathbf{g} \right) \vert w_{i,\mathbf{g}}  \rangle
\end{eqnarray}

Then, inserting Eq. (\ref{eqn:rep_nabla}) into Eq. (\ref{eqn:MLC_long}) provides:
\begin{eqnarray}
\label{eqn:MLCapp1}
\mathbf{M}_\text{LC} =  - \frac{1}{2c \Omega} \Re \sum_i^\text{occ} \sum_{\mathbf{g}} \langle w_{i,\mathbf{g}}  \vert \left( \mathbf{r} - \mathbf{g} \right)  \wedge \big[ \boldsymbol{\nabla_r} / \imath \nonumber \\
+ \imath \left( \mathbf{r} - \mathbf{g} \right) H \big] \vert w_{i,\mathbf{g}}  \rangle \nonumber \\
= - \frac{1}{2c \Omega} \Re \sum_i^\text{occ} \sum_{\mathbf{g}} \langle w_{i,\mathbf{g}}  \vert \left( \mathbf{r} - \mathbf{g} \right)  \wedge \boldsymbol{\nabla_r} / \imath \vert w_{i,\mathbf{g}}  \rangle
\end{eqnarray}
where we have used $\left( \mathbf{r} - \mathbf{g} \right)  \wedge \left( \mathbf{r} - \mathbf{g} \right) = \mathbf{0}$. 

Inserting Eq. (\ref{eqn:dkbloch_w}) into Eq. (\ref{eqn:MLCapp1}) gives:
\begin{eqnarray}
\label{eqn:MLCapp2}
\mathbf{M}_\text{LC} = - \frac{1}{2c} \Re \sum_i^\text{occ} \int_{\text{BZ}} d \mathbf{k} \ \sum_{\mathbf{g} \mathbf{g}^\prime}  \langle w_{i,\mathbf{g}^\prime}  \vert \left( \mathbf{r} - \mathbf{g}^\prime \right)  \wedge  \nonumber \\
 e^{- \imath \mathbf{k} \cdot \left( \mathbf{g}^\prime - \mathbf{g} - \mathbf{r} + \mathbf{r} \right)}  \boldsymbol{\nabla_r} / \imath    \vert w_{i,\mathbf{g}}  \rangle \nonumber \\
= -\frac{1}{2c} \Re \sum_i^\text{occ} \int_{\text{BZ}} d \mathbf{k} \ \big( -\imath \sum_{\mathbf{g}}  \langle \boldsymbol{\nabla}_\mathbf{k} u_{i,\mathbf{k}}  \vert \nonumber \\
\wedge  e^{-i \mathbf{k} \cdot \left( \mathbf{r} - \mathbf{g} \right)}  \boldsymbol{\nabla_r} / \imath    \vert w_{i,\mathbf{g}}  \rangle \big) \nonumber \\
=  \frac{1}{2c} \Re \sum_i^\text{occ} \int_{\text{BZ}} d \mathbf{k} \ \sum_{\mathbf{g}}  \langle \boldsymbol{\nabla}_\mathbf{k} u_{i,\mathbf{k}}  \vert \wedge  e^{-i \mathbf{k} \cdot \left( \mathbf{r} - \mathbf{g} \right)} \boldsymbol{\nabla_r} \vert w_{i,\mathbf{g}}  \rangle 
\end{eqnarray}
Comparing Eq. (\ref{eqn:MLCapp2}) with Eq. (\ref{eqn:MLC}), we conclude:
\begin{equation}
\label{eqn:Hu_app}
\wedge H_\mathbf{k} / \imath \vert \boldsymbol{\nabla_k} u_{i,\mathbf{k}} \rangle = \wedge \sum_\mathbf{g} e^{-\imath \mathbf{k} \cdot \left( \mathbf{r} - \mathbf{g} \right)} \boldsymbol{\nabla_r} \vert w_{i,\mathbf{g}} \rangle + a \vert \boldsymbol{\nabla}_\mathbf{k} u_{i,\mathbf{k}}  \rangle
\end{equation}
where, we have used, for arbitrary vectors $\mathbf{A}$ $\mathbf{B}$, $\mathbf{C}$ and scalar $a$, $\mathbf{A} \wedge \mathbf{B} = \mathbf{A} \wedge \mathbf{C} \to \mathbf{C}=\mathbf{B}+a\mathbf{A}$. Inserting Eq. (\ref{eqn:Hu_app}) into Eq. (\ref{eqn:MLCapp2}), and using the zero Chern invariant condition for band insulators of Eq. (\ref{eqn:chern_zero}), we find:
\begin{eqnarray}
\label{eqn:MLCapp3}
\mathbf{M}_\text{LC} &=& \Re \sum_i^\text{occ} \int_{\text{BZ}} d \mathbf{k} \ \langle \boldsymbol{\nabla}_\mathbf{k} u_{i,\mathbf{k}}  \vert \wedge \sum_\mathbf{g} e^{-\imath \mathbf{k} \cdot \left( \mathbf{r} - \mathbf{g} \right)} \boldsymbol{\nabla_r} \vert w_{i,\mathbf{g}} \rangle \nonumber \\
&+& a \times \mathbf{0}
\end{eqnarray} 
which shows that $\mathbf{M}_\text{LC}$ is invariant to the particular choice of $a$. We make the simplest possible choice $a=0$, and obtain, by comparing Eq. (\ref{eqn:MLCapp3}) with Eq. (\ref{eqn:MLC}):
\begin{eqnarray}
\Re \sum_i^\text{occ} \int_{\text{BZ}} d \mathbf{k} \ \langle \boldsymbol{\nabla}_\mathbf{k} u_{i,\mathbf{k}}  \vert \wedge \sum_\mathbf{g} e^{-\imath \mathbf{k} \cdot \left( \mathbf{r} - \mathbf{g} \right)} \boldsymbol{\nabla_r} \vert w_{i,\mathbf{g}} \rangle \nonumber \\
= \Re \sum_i^\text{occ} \int_{\text{BZ}} d \mathbf{k} \ \langle \boldsymbol{\nabla_k} u_{i,\mathbf{k}} \vert \wedge H_\mathbf{k} / \imath \vert \boldsymbol{\nabla_k} u_{i,\mathbf{k}} \rangle
\end{eqnarray}
Hence:
\begin{equation}
\wedge H_\mathbf{k} / \imath \vert \boldsymbol{\nabla_k} u_{i,\mathbf{k}} \rangle \to \wedge \sum_\mathbf{g} e^{-\imath \mathbf{k} \cdot \left( \mathbf{r} - \mathbf{g} \right)} \boldsymbol{\nabla_r} \vert w_{i,\mathbf{g}} \rangle
\end{equation}
which is non-other than the first statement in Eq. (\ref{eqn:u_Hgradk1}). The second equality in Eq. (\ref{eqn:u_Hgradk1}) is then directly obtained from Eq. (\ref{eqn:drbloch_w}).

\section{Momentum Conservation}
\label{app:mom}
The matrix-elements in Eq. (\ref{eqn:lambda_herm1}) have the form:
\begin{equation}
\int_{-\mathbf{\infty}}^{\mathbf{\infty}} d \mathbf{r} \ \psi_{i,\mathbf{k}}^\ast \left( \mathbf{r} \right) e^{\imath \mathbf{q} \cdot \mathbf{r}} \hat{O} \left( \mathbf{r} \right) \psi_{j,\mathbf{k}^\prime} \left( \mathbf{r} \right)
\end{equation}
where $\hat{O} \left( \mathbf{r} \right)$ is a periodic operator $\hat{O} \left( \mathbf{r} \right)=\hat{O} \left( \mathbf{r}-\mathbf{g} \right)$, and $\mathbf{g}=l  \mathbf{a}$ for some integer $l$ and lattice parameter $\mathbf{a}$. Writting the COs in terms of Bloch functions, we have:
\begin{eqnarray}
\int_{-\mathbf{\infty}}^{\mathbf{\infty}} d \mathbf{r} \ e^{\imath \left( \mathbf{k}^\prime - \mathbf{k} +\mathbf{q} \right) \cdot \mathbf{r}} u_{i,\mathbf{k}}^\ast \left( \mathbf{r} \right) \hat{O} \left( \mathbf{r} \right) u_{j,\mathbf{k}^\prime} \left( \mathbf{r} \right) = \nonumber \\
\sum_{l \in \mathbb{Z}} e^{\imath \left( \mathbf{k}^\prime - \mathbf{k} +\mathbf{q} \right) \cdot l \mathbf{a}} \int_{\mathbf{0}}^{\mathbf{a}} d \mathbf{r} \ \ e^{\imath \left( \mathbf{k}^\prime - \mathbf{k} +\mathbf{q} \right) \cdot \mathbf{r}} u_{i,\mathbf{k}}^\ast \left( \mathbf{r} \right) \hat{O} \left( \mathbf{r} \right) u_{j,\mathbf{k}^\prime} \left( \mathbf{r} \right)
\end{eqnarray}
where we have used the lattice periodicity of $\hat{O}$ and $u$. Using the identity:
\begin{equation}
\sum_{l \in \mathbb{Z}} e^{\imath \left( \mathbf{k}^\prime - \mathbf{k} +\mathbf{q} \right) \cdot l \mathbf{a}} = \frac{2 \pi}{\mathbf{a}} \cdot \delta( \mathbf{k}^\prime - \mathbf{k} +\mathbf{q}  )
\end{equation}
the integral gives zero unless $\mathbf{k}^\prime - \mathbf{k}=-\mathbf{q}$.

\section{Magnetization with Semi-Local Functionals in Terms of a ``Band Dispersion'' Formula}
\label{app:band_disp}
Following Eqs. (\ref{eqn:MLC}), (\ref{eqn:MIC}), (\ref{eqn:MLC3}) and (\ref{eqn:MIC3}), we may write the orbital magnetization as:
\begin{eqnarray}
\label{MtotLC}
\mathbf{M} &=& \frac{1}{c} \Re \int_\text{BZ} d \mathbf{k} \sum_i^\text{occ} \langle \boldsymbol{\nabla_k} u_{i,\mathbf{k}} \vert \wedge H_{\mathbf{k}} / \imath  \vert \boldsymbol{\nabla_k} u_{i,\mathbf{k}} \rangle \nonumber \\
&=& \frac{1}{c} \Re \int_\text{BZ} d \mathbf{k} \sum_i^\text{occ} \Big[ \langle \boldsymbol{\nabla_k} u_{i,\mathbf{k}} \vert \wedge H_{\mathbf{k}} / \imath  \vert \boldsymbol{\nabla_k} u_{i,\mathbf{k}} \rangle \nonumber \\ 
&+& \langle \boldsymbol{\nabla_k} u_{i,\mathbf{k}} \vert \wedge \left( \boldsymbol{\nabla_k}  H_\mathbf{k} \right) / \imath \vert u_{i,\mathbf{k}} \rangle \nonumber \\
&-& \langle \boldsymbol{\nabla_k} u_{i,\mathbf{k}} \vert \wedge \left( \boldsymbol{\nabla_k}  H_\mathbf{k} \right) / \imath \vert u_{i,\mathbf{k}} \rangle \Big] 
\end{eqnarray}
Then, using $\boldsymbol{\nabla_k} \left(  H_\mathbf{k}  \vert u_{i,\mathbf{k}} \rangle \right)=\left( \boldsymbol{\nabla_k} H_\mathbf{k}  \right) \vert u_{i,\mathbf{k}} \rangle + H_\mathbf{k} \vert \boldsymbol{\nabla_k}  u_{i,\mathbf{k}} \rangle$, Eq. (\ref{eqn:M}) may be rewritten as:
\begin{eqnarray}
\label{eqn:MtotLC0}
\mathbf{M} &=& \frac{1}{c} \Re \int_\text{BZ} d \mathbf{k} \sum_i^\text{occ} \Big[ \langle \boldsymbol{\nabla_k} u_{i,\mathbf{k}} \vert \wedge \boldsymbol{\nabla_k} \Big(  H_\mathbf{k} / \imath \vert u_{i,\mathbf{k}} \rangle \Big) \nonumber \\
&-& \langle \boldsymbol{\nabla_k} u_{i,\mathbf{k}} \vert \wedge \left( \boldsymbol{\nabla_k}  H_\mathbf{k} \right) / \imath \vert u_{i,\mathbf{k}} \rangle \Big]
\end{eqnarray}
For arbitrary scalar fields $\vartheta$ and $\varphi$, inserting $\boldsymbol{\nabla} \vartheta \wedge \boldsymbol{\nabla} \varphi = - \boldsymbol{\nabla} \wedge \left( \left[ \boldsymbol{\nabla}  \vartheta \right] \varphi \right)$ into Eq. (\ref{eqn:MtotLC0}), we get:
\begin{eqnarray}
\label{eqn:MtotLC1}
\mathbf{M} &=& \frac{1}{c} \Re \int_\text{BZ} d \mathbf{k} \sum_i^\text{occ}  \Big[ - \boldsymbol{\nabla_k} \wedge \langle \boldsymbol{\nabla_k}  u_{i,\mathbf{k}} \vert  H_\mathbf{k} / \imath \vert u_{i,\mathbf{k}} \rangle  \nonumber \\
&-& \langle \boldsymbol{\nabla_k}  u_{i,\mathbf{k}} \vert \wedge \left( \boldsymbol{\nabla_k} H_\mathbf{k} / \imath  \right) \vert u_{i,\mathbf{k}} \rangle  \Big] \nonumber \\
&=& \frac{1}{c} \Re \int_\text{BZ} d \mathbf{k} \sum_i^\text{occ} \Big[ \boldsymbol{\nabla_k} \wedge \langle u_{i,\mathbf{k}} \vert  H_\mathbf{k} / \imath \vert \boldsymbol{\nabla_k}  u_{i,\mathbf{k}} \rangle^\ast  \nonumber \\
&-& \langle \boldsymbol{\nabla_k}  u_{i,\mathbf{k}} \vert \wedge \left( \boldsymbol{\nabla_k} H_\mathbf{k} / \imath  \right) \vert u_{i,\mathbf{k}} \rangle  \Big] \nonumber \\
&=& \frac{1}{c} \Re \int_\text{BZ} d \mathbf{k} \sum_i^\text{occ}  \Big[ \boldsymbol{\nabla_k} \wedge \langle u_{i,\mathbf{k}} \vert  H_\mathbf{k} / \imath \vert \boldsymbol{\nabla_k}  u_{i,\mathbf{k}} \rangle \nonumber \\
&-& \langle \boldsymbol{\nabla_k}  u_{i,\mathbf{k}} \vert \wedge \left( \boldsymbol{\nabla_k} H_\mathbf{k} / \imath  \right) \vert u_{i,\mathbf{k}} \rangle  \Big]
\end{eqnarray}
We now use the following identity obtained as a corrolary to the divergence theorem:
\begin{equation}
\label{eqn:stokes}
\iiint_V d V \ \boldsymbol{\nabla} \wedge  \mathbf{F}  = \oiint_S d \mathcal{S} \ \mathbf{n}_k \wedge  \mathbf{F} 
\end{equation}
For some vector field $\mathbf{F}$ over the volume $V$ bounded by the closed surface $\mathcal{S}$ with vector normal $\mathbf{n}_k$.
Inserting Eq. (\ref{eqn:stokes}) into Eq. (\ref{eqn:MtotLC1}) we get:
\begin{eqnarray}
\label{eqn:MtotLC2}
\mathbf{M} &=& \frac{1}{c} \Re \sum_i^\text{occ} \Big[ \oiint_S d\mathcal{S} \ \mathbf{n}_k  \wedge \langle u_{i,\mathbf{k}} \vert  H_\mathbf{k} / \imath \vert \boldsymbol{\nabla_k} u_{i,\mathbf{k}} \rangle \nonumber \\
&-& \int_{\text{BZ}} d \mathbf{k} \ \langle \boldsymbol{\nabla_k}  u_{i,\mathbf{k}} \vert \wedge \left( \boldsymbol{\nabla_k} H_\mathbf{k} / \imath  \right)  \vert u_{i,\mathbf{k}} \rangle  \Big]
\end{eqnarray}
Over $\mathcal{S}$ (the edge of the Brillouin zone, where $e^{\imath \mathbf{k} \cdot \mathbf{g}}$ for lattice vector $\mathbf{g}$ is pure real) the real part of the integrand in the first term vanishes because $\psi_{i,\mathbf{k}}$ is pure real. We now use:
\begin{eqnarray}
\label{eqn:hkun}
\left( \boldsymbol{\nabla_k} H_\mathbf{k}  \right) \vert u_{i,\mathbf{k}} \rangle &=& \boldsymbol{\nabla_k} \vert H_\mathbf{k} u_{i,\mathbf{k}} \rangle - H_\mathbf{k} \vert \boldsymbol{\nabla_k} u_{i,\mathbf{k}} \rangle \nonumber \\
&=& \left( \boldsymbol{\nabla_k} \epsilon_{i,\mathbf{k}}  \right) \vert u_{i,\mathbf{k}} \rangle + \epsilon_{i,\mathbf{k}} \vert \boldsymbol{\nabla_k}  u_{i,\mathbf{k}} \rangle \nonumber \\
&-& H_\mathbf{k} \vert \boldsymbol{\nabla_k}  u_{i,\mathbf{k}} \rangle 
\end{eqnarray}
Inserting Eq. (\ref{eqn:hkun}), as well as the identity $\sum_l^\text{all} \vert u_{l,\mathbf{k}}  \rangle \langle u_{l,\mathbf{k}} \vert = \mathbbm{1}$ into Eq. (\ref{eqn:MtotLC2}) gives:
\begin{eqnarray}
\label{eqn:MtotLC3}
\mathbf{M} = - \frac{1}{c} \Re \sum_i^\text{occ} \sum_l^\text{all} \int_{\text{BZ}} d \mathbf{k} \ \Big\{ \langle \boldsymbol{\nabla_k}  u_{i,\mathbf{k}} \vert u_{l,\mathbf{k}} \rangle \wedge \nonumber \\
\Big( \langle u_{l,\mathbf{k}} \vert \left( \boldsymbol{\nabla_k} \epsilon_{i,\mathbf{k}} / \imath  \right)  \vert u_{i,\mathbf{k}} \rangle + \left(  \epsilon_{i,\mathbf{k}} - \epsilon_{l,\mathbf{k}} \right) / \imath \langle u_{l,\mathbf{k}} \vert \boldsymbol{\nabla_k}  u_{i,\mathbf{k}} \rangle  \Big) \Big\} 
\end{eqnarray}
Now, using the fact that the real part of the term with $\langle \boldsymbol{\nabla_k}  u_{i,\mathbf{k}} \vert u_{l,\mathbf{k}} \rangle \wedge \langle u_{l,\mathbf{k}} \vert \boldsymbol{\nabla_k}  u_{i,\mathbf{k}} \rangle$, being the cross product of a vector with itself vanishes, we get:
\begin{equation}
\label{eqn:MtotLC4}
\mathbf{M} = \frac{1}{c} \Re \int_\text{BZ} d \mathbf{k} \sum_i^\text{occ} \langle \boldsymbol{\nabla_k}  u_{i,\mathbf{k}} \vert u_{i,\mathbf{k}} \rangle \wedge \left( \boldsymbol{\nabla_k} \epsilon_{i,\mathbf{k}} / \imath  \right) 
\end{equation}
which is Wang and Yan's ``Band Dispersion'' term.\cite{wang2022optical}

\end{document}